\documentclass{article}

\usepackage{graphicx}
\usepackage{amsmath,amssymb,amsthm,mathtools}
\theoremstyle{plain}

\usepackage{hyperref} 
\usepackage{url}
\usepackage{multirow} 

\usepackage{listings}
\usepackage{amsmath}
\usepackage{graphicx}
\usepackage{mdframed}
\usepackage{framed}
\usepackage{wrapfig}
\usepackage{subcaption}
\usepackage{etoolbox}
\usepackage{enumitem} 
\usepackage{colortbl}
\usepackage{multirow}
\usepackage{alltt} 
\usepackage{tikz}
\usetikzlibrary{calc,arrows.meta,backgrounds,fit}
\usepackage{makecell}

\newcommand{\domainHypothesisGeneration}{\emph{Creative Hypothesis Generation}}
\newcommand{\domainStoryWriting}{\emph{Creative Story Writing}}

\newcommand{\task}{T}

\newcommand{\funcOneParam}{\mathcal{F}_0}
\newcommand{\funcTwoParam}{\mathcal{F}_1}
\newcommand{\funcThreeParam}{\mathcal{F}_2}

\newcommand{\metricDiversity}{\mathtt{CDiv}}
\newcommand{\metricEffort}{\mathtt{Effort}}
\newcommand{\metricQuality}{\mathtt{Quality}}

\newcommand{\user}{{{H}}}
\newcommand{\userset}{\mathcal{{H}}}

\newcommand{\agent}{{A}}

\newcommand{\artifact}{{{O}}}
\newcommand{\artifactset}{\mathcal{O}}

\newcommand{\adoptionRate}{{{x}}}
\newcommand{\condHumanOnly}{\textsc{HHH:All}}
\newcommand{\condAIOnly}{\textsc{AAA:Copy}}
\newcommand{\condAIDivConHumanOut}{\textsc{AAH:Finalize}}
\newcommand{\condHumanDivConAIOut}{\textsc{HHA:Draft}}
\newcommand{\condHumanDivAIConvOut}{\textsc{HAA:HIdeas}}
\newcommand{\condAIDivHumanConvOut}{\textsc{AHH:AIdeas}}
\newcommand{\condAIOnlyLow}{\textsc{AAA:Copy:One}}
\newcommand{\condAIDivConHumanOutLow}{\textsc{AAH:Finalize:One}}
\newcommand{\condHumanDivConAIOutLow}{\textsc{HHA:Draft:One}}
\newcommand{\condHumanDivAIConvOutLow}{\textsc{HAA:HIdeas:One}}
\newcommand{\condAIDivHumanConvOutLow}{\textsc{AHH:AIdeas:One}}
\newcommand{\condAIOnlyHigh}{\textsc{AAA:Copy:Mix}}
\newcommand{\condAIDivConHumanOutHigh}{\textsc{AAH:Finalize:Mix}}
\newcommand{\condHumanDivConAIOutHigh}{\textsc{HHA:Draft:Mix}}
\newcommand{\condHumanDivAIConvOutHigh}{\textsc{HAA:HIdeas:Mix}}
\newcommand{\condAIDivHumanConvOutHigh}{\textsc{AHH:AIdeas:Mix}}

\newcommand{\promptinput}[1]{{\textcolor{promptinputcolor}{{#1}}}}

\definecolor{humanAllColor}{HTML}{C8C8C8}
\definecolor{aiPolishingColor}{HTML}{A5D6A7}
\definecolor{aiAllColor}{HTML}{EF9A9A}
\definecolor{humanPolishingColor}{HTML}{90CAF9}
\definecolor{turquoiseCell}{HTML}{80DEEA}
\definecolor{springgreenCell}{HTML}{FFFFCC}
\definecolor{promptinputcolor}{rgb}{0,0,2.55}

\usepackage[main, final]{neurips_2026}

\usepackage[utf8]{inputenc} 
\usepackage[T1]{fontenc}    
\usepackage{hyperref}       
\usepackage{url}            
\usepackage{booktabs}       
\usepackage{amsfonts}       
\usepackage{nicefrac}       
\usepackage{microtype}      
\usepackage{xcolor}         

\title{Dynamics of Collective Diversity in\\ Human-AI Co-Creation for Creative Tasks}

\author{%
  Manh Hung Nguyen\\
  MPI-SWS, Germany\\
  \text{manguyen@mpi-sws.org} \\
  \And
  Chao Wen\\
  MPI-SWS, Germany\\
  \text{chaowen@mpi-sws.org} \\
  \AND
  Sebastian Tschiatschek \\
  University of Vienna, Austria \\
  \text{sebastian.tschiatschek@univie.ac.at} \\
  \And
  Adish Singla \ \ \ \ \ \ \ \ \ \ \\
  MPI-SWS, Germany \ \ \ \ \ \ \ \ \ \ \\
  \text{adishs@mpi-sws.org} \ \ \ \ \ \ \ \ \ \ \\
}

\setcitestyle{numbers,square,comma}

\begin{document}

\maketitle

\begin{abstract}
Generative AI systems are increasingly deployed at scale to support human users in creative tasks across a variety of domains. However, as more users adopt the same AI agent, the resulting artifacts may become increasingly homogeneous, leading to a collapse in collective diversity. Prior work has documented this risk, but lacks a systematic framework for understanding how collective diversity changes as AI adoption scales across co-creation settings. In this work, we conduct a large-scale crowdsourcing study to understand and model the dynamics of collective diversity, quality, and effort in human-AI co-creation across two creative domains: \domainHypothesisGeneration{} and \domainStoryWriting{}. We find that current AI agents used in co-creation can improve artifact quality and reduce effort, but at the cost of declining collective diversity as adoption increases. We further show that more fine-grained human-AI interaction, along with more diverse agent policies, can mitigate this collapse. Finally, we fit predictive models that quantify how collective diversity scales with adoption across co-creation settings, providing insights for designing AI agents that aim to preserve collective diversity.
\end{abstract}

\section{Introduction} \label{sec:introduction}

Generative AI systems are increasingly deployed at scale to support creative work across domains such as design, science, and education~\cite{DBLP:journals/corr/abs-2609-07711,DBLP:journals/corr/abs-2512-18388,DBLP:journals/nature/WangFD0HLCLKDAB23,kasneci2023chatgpt}. In many such settings, large populations of users interact with the same AI agent (e.g., ChatGPT) to co-create artifacts such as written content, product designs, and scientific ideas~\cite{chatterji2025people}. While adopting AI agents can improve artifact quality and reduce effort for individual users~\cite{noy2023experimental,dell2023navigating,DBLP:journals/corr/abs-2302-06590}, widespread adoption of the same agent may also make independently produced artifacts more homogeneous at the population level, limiting the breadth of ideas and solutions generated by the population~\cite{doshi2024generative,moon2025homogenizing}.
This motivates our central research question:
\emph{How does collective diversity scale as AI agents are adopted across a population, and how can human-AI co-creation systems be designed to preserve collective diversity?}

Recent work has begun to document the risk of diversity collapse, showing that generative AI can improve individual outputs while making outputs across users more similar~\cite{doshi2024generative,DBLP:conf/candc/AndersonSK24}.
Follow-up studies have characterized related effects, such as slower diversity growth~\cite{moon2025homogenizing} and persistent homogenization after AI assistance is withdrawn~\cite{ZHOU2026103087}.
However, these studies typically examine a limited number of configurations of agents, tasks, and user populations. As a result, it remains a challenge to have a good understanding of how collective diversity changes as AI adoption increases across a population, how agent design choices can mitigate diversity collapse, and how to predict the dynamics of collective diversity for new deployment settings, such as a different agent policy or interaction mode.

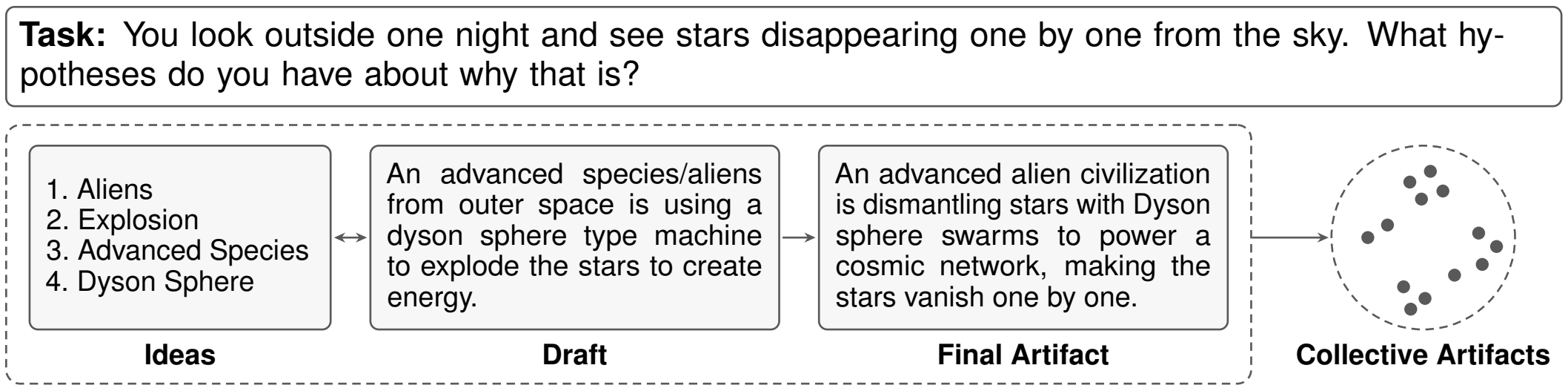
\begin{figure}[t]
\centering
\colorlet{cardc}{black!65}
\newcommand{\cardh}{2.3cm}%
\tikzset{
  tbox/.style={draw=cardc, line width=0.7pt, rounded corners=3pt, fill=black!3, minimum height=\cardh, align=justify, font=\normalsize, inner xsep=6pt, inner ysep=5pt, anchor=north},
  clab/.style={font=\normalsize\bfseries, anchor=base},
  flow/.style={-{Stealth[length=4pt, width=3.5pt]}, cardc, line width=0.7pt},
}
\newcommand{\card}[5]{
  \node[tbox, text width=#3] (#5) at (#1,#2) {\hyphenpenalty=10000\exhyphenpenalty=10000 #4\par};%
}
\resizebox{\linewidth}{!}{\sffamily
\begin{tikzpicture}
\card{1.885}{3.35}{3.35cm}{%
  1.~Aliens \\
  2.~Explosion \\
  3.~Advanced Species \\
  4.~Dyson Sphere}{I}
\card{6.83}{3.35}{4.7cm}{An advanced species/\allowbreak aliens from outer space is using a dyson sphere type machine to explode the stars to create energy.}{D}
\draw[flow, {Stealth[length=4pt, width=3.5pt]}-{Stealth[length=4pt, width=3.5pt]}] ($(I.east)+(0.03,0)$)--($(D.west)+(-0.03,0)$);
\node[clab] (Ic) at ($(I.south)+(0,-0.42)$) {Ideas};
\node[clab] (Dc) at ($(D.south)+(0,-0.42)$) {Draft};
\card{12.45}{3.35}{4.7cm}{An advanced alien civilization is dismantling stars with Dyson sphere swarms to power a cosmic network, making the stars vanish one by one.}{F}
\draw[flow] ($(D.east)+(0.03,0)$)--($(F.west)+(-0.03,0)$);
\node[clab] (Fc) at ($(F.south)+(0,-0.42)$) {Final Artifact};
\begin{scope}[on background layer]
  \node[draw=cardc, line width=0.7pt, dash pattern=on 3pt off 2pt, rounded corners=5pt, inner xsep=8pt, inner ysep=7pt,
        fit=(I.south west)(F.north east)(Ic.base)(Dc.base)(Fc.base)] (G) {};
\end{scope}
\path let \p1=(D.north west), \p2=(D.south east) in
  node[circle, draw=cardc, line width=0.7pt, dash pattern=on 3pt off 2pt, inner sep=0, outer sep=0,
       minimum size=\y1-\y2-0.7pt] (C) at (17.46,{(\y1+\y2)/2}) {};
\foreach \px/\py in {-0.15/0.62, 0.08/0.74, -0.02/0.43, 0.22/0.52,
                      0.62/0.05, 0.82/-0.1, 0.66/-0.3,
                      -0.22/-0.55, 0.02/-0.68, -0.14/-0.8,
                      -0.4/0.14, 0.35/-0.42, -0.62/0}{
  \fill[cardc] ($(C.center)+(1.12*\px,1.12*\py)$) circle (2.3pt);
}
\node[clab] (Cc) at (C.center |- Fc.base) {Collective Artifacts};
\draw[flow, shorten >=1pt] (C.west -| G.east)--(C.west);
\path let \p1=(G.north west), \p2=(Cc.east) in
  node[draw=cardc, line width=0.7pt, rounded corners=3pt, font=\large, align=left, inner sep=5pt, anchor=south west,
       text width=\x2-\x1-10pt] (T) at (\x1,\y1+6pt)
 {\textbf{Task:} You look outside one night and see stars disappearing one by one from the sky. What hypotheses do you have about why that is?};
\end{tikzpicture}
}
    \caption{Human-AI co-creation at the population level, illustrated through \domainHypothesisGeneration{}. We study the collective diversity of artifacts produced across users given the same task under different co-creation settings.}
    \label{fig:illustrative-example}
\end{figure}

In this paper, we study and model the population-level dynamics of collective diversity in human-AI co-creation under increasing adoption of generative AI agents.
To this end, we conduct a large-scale crowdsourcing study across two creative domains, \domainHypothesisGeneration{} and \domainStoryWriting{}, collecting artifacts from various human-AI co-creation settings (see Figure~\ref{fig:illustrative-example}). Our main results and contributions are summarized as follows:
\begin{itemize}[leftmargin=*]
    \item First, we analyze collective diversity under deployment settings where all adopting users co-create with the same AI agent. We find that AI assistance improves individual artifact quality and reduces effort, but at the cost of declining collective diversity as adoption increases (Section~\ref{sec:evaluation1}).

    \item Second, we examine richer co-creation settings to mitigate diversity collapse. We show that more fine-grained human-AI interaction, together with more diverse agent policies, can help preserve collective diversity (Section~\ref{sec:evaluation2}). 

    \item Third, we fit predictive models that estimate how collective diversity scales with adoption across co-creation settings. This enables us to predict diversity under new deployment settings, informing the design of AI agents that better preserve collective diversity (Section~\ref{sec:evaluation3}).
\end{itemize}
\looseness-1We release our implementation and resources for analysis to support future research on human-AI co-creation systems.\footnote{GitHub repo: \url{https://github.com/machine-teaching-group/neurips2026-collective-diversity}}

\section{Related Work} \label{sec:related-work}

\looseness-1
\textbf{Generative Models for Content Generation.}
Generative models have been increasingly used for content generation across various types of artifacts, such as code, images, and videos. Prior works have primarily focused on tasks centered around boosting productivity or achieving higher accuracy, such as code generation~\cite{DBLP:journals/corr/abs-2107-03374,DBLP:journals/corr/abs-2308-12950,DBLP:conf/acl/WenSS25} and mathematical reasoning~\cite{DBLP:journals/corr/abs-2402-03300,DBLP:conf/iclr/LightmanKBEBLLS24}. 
More recent work has explored creative domains, such as story writing~\cite{DBLP:conf/emnlp/TelekiBDJLLWLZSC25,DBLP:journals/corr/abs-2401-17268}, educational task generation~\cite{2026.EDM.poster-demo-papers.424,DBLP:conf/aied/NguyenPGTS25}, and scientific discovery~\cite{DBLP:journals/corr/abs-2408-06292,DBLP:journals/corr/abs-2502-18864}, where novelty and diversity are more emphasized than just accuracy. 
In creative domains, however, generative models have been found to exhibit an ``Artificial Hivemind'' effect, i.e., models produce homogeneous outputs both within a single model and across different models~\cite{DBLP:conf/nips/JiangCLLFDTSC25,DBLP:conf/iclr/WuBC25}. To mitigate this, existing work has proposed improving diversity through different strategies, such as prompting-based strategies that explicitly encourage exploration of diverse ideas~\cite{2026.EDM.poster-demo-papers.424}, decoding-time strategies that adjust the token sampling distribution~\cite{DBLP:conf/iclr/MinhBNR0S25}, and training-time strategies that reward diversity~\cite{modified_DBLP:journals/corr/abs-2505-23433,DBLP:journals/corr/abs-2503-17126}.
However, these works focus on model-only settings, studying or optimizing diversity at the level of isolated models and assuming no human involvement. Instead, we study human-AI co-creation and focus on collective diversity at the population level, rather than solely on model outputs.

\looseness-1
\textbf{Human-AI Co-Creation.}
Prior work has studied human-AI co-creation from different perspectives, including controlled experiments that examine the impact of generative models on humans during the co-creation process~\cite{DBLP:conf/chi/KumarVJA25,DBLP:conf/chi/Wadinambiarachchi24}, as well as the design of interfaces to support co-creation~\cite{DBLP:journals/corr/abs-2512-18388,DBLP:conf/chi/SuhCMLX24,DBLP:conf/chi/WangLC0C25,DBLP:conf/chi/ZhouLTT0CW24}.
However, most of these works focus on individual-level outcomes, such as user experience, creativity support, user agency, and artifact quality~\cite{DBLP:journals/corr/abs-2512-18388,DBLP:conf/chi/SuhCMLX24,DBLP:conf/chi/WangLC0C25,DBLP:conf/chi/ZhouLTT0CW24}. More recent studies have begun to examine human-AI co-creation at the population level, finding that while generative AI can improve individual outcomes, it may also lead to diversity collapse~\cite{doshi2024generative,DBLP:conf/candc/AndersonSK24}. Follow-up work has further characterized related diversity collapse effects by proposing the metric of diversity growth rate~\cite{moon2025homogenizing}.
Despite these insights into collective diversity, existing studies are restricted to a small number of fixed configurations of agents, tasks, and user populations, and primarily show that collective diversity can be undermined. Moreover, they do not systematically investigate how the dynamics of collective diversity can be predicted under different design choices.
Our work bridges these gaps by conducting large-scale studies to understand how collective diversity evolves, examine what design choices can preserve collective diversity, and develop predictive models of diversity dynamics under different settings.

\section{Problem Setup}
\label{sec:problem-setup}

\textbf{Preliminaries.} Let $\task$ denote a creative task with the objective of producing an artifact $\artifact$. Let $\user$ denote a human user who works on this task, either alone or along with an AI agent. An AI agent $\agent$ is characterized by a generative policy and an interaction mode. The policy can be a single policy or a mixture of policies (e.g., over different models and personas); the interaction mode specifies how the agent interacts with the user. Each agent characterizes a unique co-creation setting.
We denote the outputs of a human-AI co-creation session as $\artifact^{\task}_{\user, \agent}$ (a produced artifact) and $E^{\task}_{\user, \agent}$ (additional meta-information, e.g., time taken while producing the artifact).
In cases where the human works on the task alone, we denote the outputs as $\artifact^{\task}_{\user, \text{none}}$ and $E^{\task}_{\user, \text{none}}$.

\textbf{Co-Creation from a User's Perspective.} Given a task $\task$ and the option to work with an agent $\agent$, we consider the common case where a user $\user$ primarily aims to produce a high-quality artifact with minimal effort. Let $\metricQuality(\artifact; \task)$ denote the quality of artifact $\artifact$ and let $\metricEffort(E^{\task}_{\user, \agent}; \task)$ denote the associated effort. From a user's perspective, quality gains and effort savings may encourage them to co-create with agent $\agent$ instead of working alone.

\textbf{Co-Creation from the System's Perspective.} In practice, the agent $\agent$ is not deployed to a single user, but rolled out across a population of users $\userset$. In this setting, each user $\user \in \userset$ works on the same task and produces an artifact, yielding a collection of artifacts $\artifactset^{\task}_{\userset, \agent} = \{\artifact^{\task}_{\user, \agent} \mid \user \in \userset\}$ along with $\mathcal{E}^{\task}_{\userset, \agent} = \{E^{\task}_{\user, \agent} \mid \user \in \userset\}$. On the one hand, the system seeks to improve the average quality of artifacts denoted as $\metricQuality(\artifactset^{\task}_{\userset, \agent}; \task)$  and reduce the average effort denoted as $\metricEffort(\mathcal{E}^{\task}_{\userset, \agent}; \task)$. On the other hand, the system aims to maximize the collective diversity of artifacts produced by the population associated with this creative task. We measure this collective diversity as $\metricDiversity(\artifactset^{\task}_{\userset, \agent})$.

\looseness-1\textbf{Research Questions (RQs).} Prior work suggests that working with AI agents can improve individual outcomes while reducing the diversity of artifacts produced by a population~\cite{doshi2024generative,moon2025homogenizing}. 
More concretely, let $\adoptionRate \in [0,1]$ denote the fraction of users who adopt the agent and let $d=(\agent,\adoptionRate)$ denote a deployment setting.
We seek to understand how the system could achieve a desired collective diversity of artifacts through the design of agent $\agent$ (i.e., its policy and interaction mode) or by directly controlling the adoption rate $\adoptionRate$. 
We investigate the following three research questions: {(RQ1)} Under a typical deployment setting, where adopting users co-create with the same agent that uses a single policy and a simple interaction mode, how do collective diversity, average quality, and average effort scale with the adoption rate? (Section~\ref{sec:evaluation1}); {(RQ2)} How do these population-level dynamics change under richer deployment settings, where the agent uses a mixture of policies or a more fine-grained interaction mode that lets users iterate on ideas with the agent? (Section~\ref{sec:evaluation2}); {(RQ3)} Can we predict how collective diversity scales with the adoption rate for a new co-creation setting before deployment? (Section~\ref{sec:evaluation3}).

\section{RQ1: Collective Diversity in Typical Deployment Settings}
\label{sec:evaluation1}

\subsection{Definitions and Formulations}
\label{subsec:evaluation1-formulation}

\textbf{Definitions for Collective Diversity.}  We measure the diversity of a set of artifacts $\mathcal{O}$ as $\metricDiversity(\mathcal{O}):=\mathbb{E}_{o \sim \mathcal{O}, o' \sim \mathcal{O}} \left[1-k(o,o')\right]$, where $k(\cdot,\cdot)$ is a kernel function measuring the similarity between two artifacts~\cite{rao1982diversity}. For example, $k(o,o')$ could be a Gaussian kernel based on embeddings of artifacts~\cite{murphy2012machine}; for the special case of $k(o,o')$ being an indicator function, this notion of diversity reduces to the Gini--Simpson index of diversity~\cite{simpson1949measurement}. Moreover, for two sets of artifacts $\mathcal{O}$ and $\tilde{\mathcal{O}}$, we measure a notion of dissimilarity between them as $\text{Dis}(\mathcal{O}, \tilde{\mathcal{O}}):=\mathbb{E}_{o \sim \mathcal{O}, o' \sim \tilde{\mathcal{O}}} \left[1-k(o,o')\right]$.

\textbf{Dynamics in Deployment Setting.} Consider a deployment setting $d = (\agent, \adoptionRate)$ for task $\task$ and population $\mathcal{H}$. Let $\mathcal{O}^T_{\mathcal{H}, \agent}$ be the set of artifacts produced when each user co-creates with the agent, $\mathcal{O}^T_{\mathcal{H}, \text{none}}$ be the artifacts produced when each user works without the agent, and $\mathcal{O}^T_{\text{none}, \agent}$ be the artifacts produced when the agent works alone (or equivalent to when the user simply copies the artifact produced by the agent). At adoption $\adoptionRate$, a fraction $\adoptionRate$ of users co-create with $\agent$ and the rest work alone. Using the produced artifact sets and the metric definitions, we can write the collective diversity $\metricDiversity(A, x;T)$, average quality $\metricQuality(A, x;T)$, and average effort $\metricEffort(A, x;T)$ at adoption $\adoptionRate$ equivalently as follows:

\vspace{-2mm}

{\belowdisplayskip=0pt
\begin{equation}
     \metricDiversity(A, x;T)
     =
     (1-\adoptionRate)^2  \metricDiversity(\mathcal{O}^T_{\mathcal{H}, \text{none}})
     + \adoptionRate^2  \metricDiversity(\mathcal{O}^T_{\mathcal{H}, \agent})
     + 2 (1- \adoptionRate) \adoptionRate  \text{Dis}(\mathcal{O}^T_{\mathcal{H}, \text{none}}, \mathcal{O}^T_{\mathcal{H}, \agent}),   \label{eq:diversity-adoption-quadratic}
     \vspace{-4mm}
 \end{equation}}
{
\begin{flalign}
    \ \ &\metricQuality(\agent, \adoptionRate; \task)
    =
    (1 - \adoptionRate) \metricQuality(\artifactset^{\task}_{\userset, \text{none}}; \task)
    + \adoptionRate \metricQuality(\artifactset^{\task}_{\userset, \agent}; \task),
    &&\label{eq:quality-adoption-linear}\\
    \ \ &\metricEffort(\agent, \adoptionRate; \task)
    \ \ =
    (1 - \adoptionRate) \metricEffort(\mathcal{E}^{\task}_{\userset, \text{none}}; \task)
    + \adoptionRate \metricEffort(\mathcal{E}^{\task}_{\userset, \agent}; \task).
    &&\label{eq:effort-adoption-linear}
\end{flalign}}%

In terms of functional forms, the collective diversity is a quadratic function of the adoption rate (i.e., in the form of $y(x)=a+bx+cx^2$), while the average quality and average effort are linear in the adoption rate (i.e., in the form of $y(x)=a+bx$). These functional forms will be useful later in Section~\ref{sec:evaluation3} when we learn predictive models. 

\looseness-1
\textbf{Extreme Case of Diversity Collapse.} To gain some intuition, we consider here an extreme case of $x$ close to $1$, where most users adopt the agent, copy its produced artifacts, and the collective diversity is thus driven by the agent's output diversity.
For instance, when considering typical deployment with agents like ChatGPT, high adoption rates are driven by the fact that AI agents can provide good quality artifacts with low effort from human users. In this case, when all users are working with the agent and simply copying the artifacts, the collective diversity will simply be equivalent to $\metricDiversity(\mathcal{O}^T_{\text{none}, \agent})$ which represents the diversity associated with the agent's outputs. Given the low diversity on creative outputs associated with typical agents~\cite{DBLP:conf/nips/JiangCLLFDTSC25}, the collective diversity collapses with increasing adoption in this extreme case. 

\looseness-1Beyond this case where users simply copy the agent's output, users may draft an answer for the agent to finalize, or finalize the agent's draft. Table~\ref{tab:typical-settings} summarizes the settings we study in RQ1, identifying human and AI roles in ideation (generating ideas), drafting (drafting an answer), and finalization (editing or adapting the draft into the final answer). 
Concretely, this table covers typical settings that use an agent with a single generative policy (referred to as \textsc{One}) and simple interaction modes.

\begin{table}[h]
\centering
\caption{Co-creation settings for our RQ1 study. H and A denote human and AI agent roles, respectively, in ideation, drafting, and finalization. The suffix \textsc{:One} indicates that the agent uses a single generative policy (i.e., one generative model).}
\label{tab:typical-settings}

\centering
\renewcommand{\arraystretch}{1}
\setlength{\tabcolsep}{4pt}
\newlength{\settingwd}
\settowidth{\settingwd}{%
    \begin{tabular}{@{}l@{}}
    \textbf{Setting}\\ \condHumanOnly\\ \condAIOnlyLow\\ \condAIDivConHumanOutLow\\ \condHumanDivConAIOutLow
    \end{tabular}}
{
    \begin{tabular}{@{}l@{\hskip 8pt}p{\dimexpr\columnwidth-\settingwd-8pt\relax}@{}}
    \toprule
    \textbf{Setting} & \textbf{Description} \\
    \midrule
    \condHumanOnly & Human generates ideas, drafts an answer, and finalizes
it. \\
    \condAIOnlyLow & AI agent generates ideas, drafts an answer, and finalizes it. \\
    \condAIDivConHumanOutLow & AI agent generates ideas and drafts an answer, which the human finalizes. \\
    \condHumanDivConAIOutLow & Human generates ideas and drafts an answer, which the
AI agent finalizes. \\
    \bottomrule
    \end{tabular}
}

\vspace{-1em}
\end{table}

\subsection{Evaluation Setup}
\label{subsec:evaluation1-setup}

In this section, we conduct a large-scale crowdsourcing study to empirically investigate the population-level scaling dynamics under typical co-creation settings in Table~\ref{tab:typical-settings} (see Appendix~\ref{app:domains} for details).

\textbf{Domains.} We consider the following two creative domains:

\begin{itemize}[leftmargin=*]
    \item \domainHypothesisGeneration. This domain involves generating creative scientific hypotheses. We use four tasks from~\cite{DBLP:conf/emnlp/IsmayilzadaLLPBPB25}, where each task provides a scenario and requires proposing a creative scientific hypothesis. For instance, one task instruction is ``You look outside one night and see stars disappearing one by one from the sky. What hypotheses do you have about why that is?''

    \item \domainStoryWriting. This domain involves writing short creative stories. We use four tasks from~\cite{DBLP:conf/icccrea/IsmayilzadaSP25}, where each task requires writing a 3--5 sentence story using three given words as creatively as possible. For instance, one task instruction is ``Write a 3–5 sentence story using: stamp, letter, and send.''
\end{itemize}

\begin{figure*}[t]
    
    \centering
    \footnotesize
    
    \begin{subfigure}[t]{\textwidth}
        
        \includegraphics[width=1.03\textwidth]{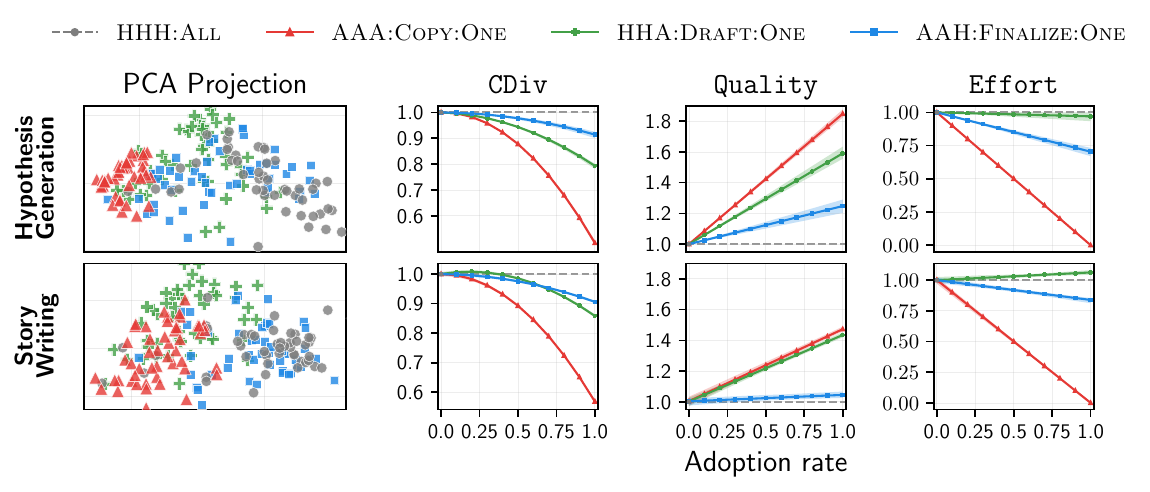}
        \vspace{-0.6cm}
        \caption{PCA projection of answers on a task, along with collective diversity, quality, and effort averaged over tasks. Metrics are normalized to the corresponding human-only baseline (\condHumanOnly). Shaded bands show $95\%$ bootstrap confidence intervals.}
        \label{fig:typical-settings-curves}
    \end{subfigure}
    
    \vspace{0.2cm}
    
     \begin{subfigure}[t]{\textwidth}
            \centering
            {%
            \renewcommand{\arraystretch}{1.2}
            \begin{tabular}{@{}c@{\hspace{6pt}}p{13.3cm}@{}}
              \toprule
              \multicolumn{2}{@{}p{13.9cm}@{}}{\textbf{Task:} You look outside one night and see stars disappearing one by one from the sky. What hypotheses do you have about why that is?} \\
              \midrule
              \colorbox{humanAllColor}{1} & The world is actually an illusion and the artist of this illusion is erasing the stars. \\
              \colorbox{humanAllColor}{2} & A massive cloud is slowly sweeping across our view of the galaxy and blocking the stars. \\
              \colorbox{humanAllColor}{3} & Aliens are sucking the energy from the stars, and therefore, the stars lose their light. \\
              \midrule
              \colorbox{aiAllColor}{1} & A vast, invisible sheet of ultra-dense dark matter is drifting across the galaxy and bending their light away one by one. \\
              \colorbox{aiAllColor}{2} & A living sheet of dark-matter-like organisms is drifting across space and selectively absorbing starlight as it feeds. \\
              \colorbox{aiAllColor}{3} & A living wave of dark matter is moving across the galaxy and temporarily cocooning each star in a light-blocking shell. \\
              \bottomrule
            \end{tabular}
            }%
         \caption{Example answers from \colorbox{humanAllColor}{\condHumanOnly} and \colorbox{aiAllColor}{\condAIOnlyLow} on a hypothesis generation task. AI-generated answers are homogeneous, causing a sharp decline in collective diversity when humans simply copy them.}
     \label{fig:typical-settings-examples}
     \end{subfigure}
     \caption{(RQ1) Results of our study in typical co-creation settings.}
\end{figure*}

\textbf{Data Collection.}
We construct an artifact dataset that contains 50 final artifacts for each task and co-creation setting (4 tasks $\times$ 4 settings $\times$ 50 = 800 artifacts per domain), either collected from participants or generated by the agent. In all settings, the agent uses \texttt{Qwen3-235B-A22B-2507-Instruct}~\cite{DBLP:journals/corr/abs-2505-09388} as the single policy.\footnote{We accessed the model via Amazon Bedrock APIs~\cite{amazon_bedrock_apis}.} Since \condAIOnlyLow{} refers to a setting where users simply copy the agent's output without editing, we generated artifacts for this setting by prompting the agent directly for each task, without human input. Then, we recruited participants from Prolific~\cite{prolific} for the settings that require human input, namely \condHumanOnly{}, \condAIDivConHumanOutLow{}, and \condHumanDivConAIOutLow{} (see Appendix~\ref{app:instructions} for the study interfaces).\footnote{\looseness-1Participants were at least 18 years old, fluent in English, and had 99--100\% approval rates. The participants were paid \pounds10/hour on average.} Each participant was assigned to one of these settings. Our studies were approved by the local institutional review board, and all participants provided informed consent before starting the study.

\looseness-1
\textbf{Evaluation Procedure.} We evaluate a given set of artifacts $\artifactset = \{\artifact_1, \ldots, \artifact_n\}$ using three metrics: diversity, average quality, and average effort. 
For diversity, we use an RBF kernel $k(\artifact,\artifact')=\exp(-\|\phi(\artifact)-\phi(\artifact')\|^2/\sigma^2)$, where $\phi(\artifact)$ is the \texttt{Qwen3-Embedding-8B}~\cite{DBLP:journals/corr/abs-2506-05176} embedding of artifact $\artifact$ and $\sigma$ is the median pairwise Euclidean distance between artifact embeddings within each domain. To evaluate the average quality, we recruited human raters from Prolific to evaluate each artifact for \textit{usability} and \textit{creativity} and map their answers to \{0-Unusable, 1-Usable\} and \{1-Low, 2-Medium, 3-High\}, respectively (see\ Appendix~\ref{app:instructions}). We define each rating's quality score as the product of usability and creativity, normalized to $[0,1]$. We average these scores within each artifact, then across artifacts. To evaluate average effort, we use completion time of the user when creating each artifact as a proxy for the effort, and we estimate average effort by averaging the completion time over all artifacts.

\subsection{Results}
\label{subsec:evaluation1-results}
Below we provide a summary of the key results, and further discuss additional results in Appendix~\ref{app:additional-experiments}.

\looseness-1\textbf{Average Quality Improves with Increasing Adoption.} Figure~\ref{fig:typical-settings-curves} shows that increasing adoption generally improves average  quality across both domains. The gains are largest under \condAIOnlyLow{}, where average quality increases most sharply and effort decreases the most. \condHumanDivConAIOutLow{} achieves higher average quality than \condAIDivConHumanOutLow{} in both domains. \condAIDivConHumanOutLow{}, however, reduces user effort, whereas effort under \condHumanDivConAIOutLow{} remains close to the \condHumanOnly{} baseline. 

\looseness-1\textbf{Collective Diversity Collapses with Increasing Adoption.} Figure~\ref{fig:typical-settings-curves} shows that diversity generally declines as adoption increases, and the decline is steepest under \condAIOnlyLow{}. Figure~\ref{fig:typical-settings-examples} shows this pattern on an example task in \domainHypothesisGeneration{}. At full adoption, the two settings where both the human and the AI agent contribute preserve more collective diversity than \condAIOnlyLow{} but still remain below \condHumanOnly{}. Overall, the typical deployment settings reveal a clear tradeoff where higher AI adoption improves average quality and can reduce user effort, but increasingly homogenizes the population's outputs.

\section{RQ2: Collective Diversity in Richer Deployment Settings}
\label{sec:evaluation2}

\subsection{Richer Co-Creation Settings}
\label{subsec:evaluation2-formulation}

Eq.~\ref{eq:diversity-adoption-quadratic} in Section~\ref{sec:evaluation1} and our empirical results highlight that the collective diversity depends on two factors: (i) how far the user and agent co-creation moves artifacts from the human-only artifacts, captured by $\text{Dis}(\mathcal{O}^T_{\mathcal{H}, \text{none}}, \mathcal{O}^T_{\mathcal{H}, \agent})$, and (ii) the diversity of the co-created artifacts, captured by $\metricDiversity(\mathcal{O}^T_{\mathcal{H}, \agent})$. In this section, we consider richer agents $\agent$ that offer more fine-grained interaction modes and use more diverse generative policies. Table~\ref{tab:richer-settings} summarizes these richer agents and the deployment settings we empirically study through a crowdsourcing study described below.

\begin{table}[h!]
\centering
\caption{Co-creation settings in our RQ2 study. H and A denote human and AI agent roles, respectively, in ideation, drafting, and finalization. The suffixes \textsc{:One} and \textsc{:Mix} indicate an agent with a single generative policy and a mixture of generative policies, respectively.}
\label{tab:richer-settings}
\renewcommand{\arraystretch}{1.4}
\setlength{\tabcolsep}{4pt}
\newlength{\richsettingwd}
\settowidth{\richsettingwd}{%
    \begin{tabular}{@{}l@{}}
    \textbf{Setting}\\ \condHumanOnly\\ \condAIOnlyLow\\ \condAIDivConHumanOutLow\\ \condHumanDivConAIOutLow\\ \condHumanDivAIConvOutLow\\ \condAIDivHumanConvOutLow
    \end{tabular}}
\begin{tabular}{@{}p{\richsettingwd}@{\hskip 8pt}p{\dimexpr\columnwidth-\richsettingwd-8pt\relax}@{}}
    \toprule
    \textbf{Setting} & \textbf{Description} \\
    \midrule
    \condHumanOnly & Same as described in Table~\ref{tab:typical-settings}. \\
    \midrule
    \condAIOnlyLow & Same as described in Table~\ref{tab:typical-settings}. \\
    \condAIDivConHumanOutLow & Same as described in Table~\ref{tab:typical-settings}. \\
    \condHumanDivConAIOutLow & Same as described in Table~\ref{tab:typical-settings}. \\
    \condHumanDivAIConvOutLow & Human generates ideas, and the AI agent drafts an answer from them and finalizes it. Human can revise the ideas and request a new draft. \\
    \condAIDivHumanConvOutLow & AI agent generates ideas, and the human drafts an answer from them and finalizes it. Human can request new ideas at any time. \\
    \midrule
    \condAIOnlyHigh & Same interaction mode as \condAIOnlyLow{}, with a mixture of policies. \\
    \condAIDivConHumanOutHigh & Same interaction mode as \condAIDivConHumanOutLow{}, with a mixture of policies. \\
    \condHumanDivConAIOutHigh & Same interaction mode as \condHumanDivConAIOutLow{}, with a mixture of policies. \\
    \condHumanDivAIConvOutHigh & Same interaction mode as \condHumanDivAIConvOutLow{}, with a mixture of policies. \\
    \condAIDivHumanConvOutHigh & Same interaction mode as \condAIDivHumanConvOutLow{}, with a mixture of policies. \\
    \bottomrule
\end{tabular}
\vspace{-1em}
\end{table}

\subsection{Evaluation Setup}
\vspace{-1mm}
\label{subsec:evaluation2-setup}

Next, we describe the evaluation setup for co-creation settings in Table~\ref{tab:richer-settings}, following the setup used earlier in Section~\ref{subsec:evaluation1-setup} (see Appendix~\ref{app:domains} for details).

\begin{figure*}[t]
    \centering
    \includegraphics[width=1\textwidth]{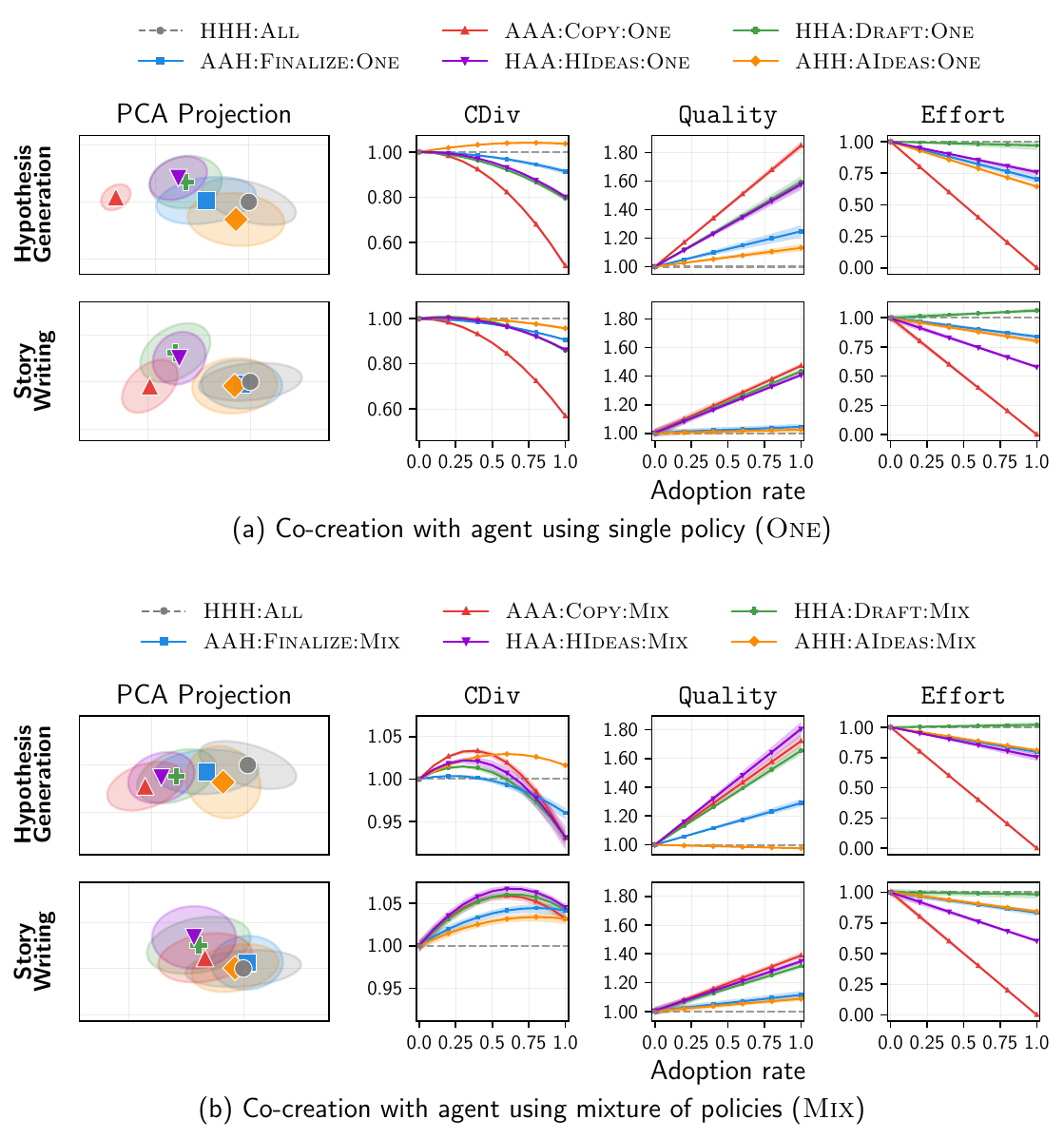}
    \caption{\looseness-1(RQ2) Collective diversity, quality, and effort tradeoffs under richer human-AI co-creation settings. PCA panels show centroids and $1\sigma$ covariance ellipses for one task per domain. Shaded bands on the curves show $95\%$ bootstrap confidence intervals. All metrics are normalized to the {\condHumanOnly}.
    }
    \label{fig:richer-settings-curves}
\end{figure*}

\looseness-1\textbf{Agent Policies.}
For each interaction mode, we compare two agent policies: (i) a single policy (\textsc{One}), supported by one model and no personas as in Section~\ref{sec:evaluation1}; and (ii) a mixture-based policy (\textsc{Mix}), which draws on multiple models and personas. In settings with mixture-based policies, given a task $\task$, the agent samples from a pool of candidate models and personas rather than relying on a single fixed model to generate all outputs. Concretely, we use a pool of $5$ models from different families and with different sizes (\texttt{Qwen3-235B-A22B-2507-Instruct}~\cite{DBLP:journals/corr/abs-2505-09388}, \texttt{DeepSeek-V3.2}~\cite{DBLP:journals/corr/abs-2512-02556}, \texttt{Kimi-K2.5}~\cite{DBLP:journals/corr/abs-2602-02276}, \texttt{gpt-oss-120b}~\cite{DBLP:journals/corr/abs-2508-10925}, and \texttt{GLM-5}~\cite{DBLP:journals/corr/abs-2602-15763}), and a pool of personas from~\cite{DBLP:journals/corr/abs-2406-20094} (see Appendix~\ref{app:prompts}). 

\textbf{Data Collection.} We use the same domains, tasks, and data collection procedures as in Section~\ref{sec:evaluation1}, with $50$ final artifacts  for each task and co-creation setting (4 tasks $\times$ 11 settings $\times$ 50 = 2200 artifacts per domain).
We reuse the data for \condHumanOnly{},  \condAIDivConHumanOutLow{}, \condHumanDivConAIOutLow{}, and \condAIOnlyLow{} from Section~\ref{sec:evaluation1}. We collect data for \condAIDivHumanConvOutLow{} and \condHumanDivAIConvOutLow{} by recruiting participants from Prolific. Next, we generate artifacts for \condAIOnlyHigh{} directly using the mixture-based agent policy, without participants. We then collect additional data for the settings with mixture-based agent policy (\condAIDivConHumanOutHigh, \condHumanDivConAIOutHigh, \condHumanDivAIConvOutHigh, and \condAIDivHumanConvOutHigh) by recruiting additional participants from Prolific.

\subsection{Results}
\label{subsec:evaluation2-results}

\textbf{Diversity Is Higher When Humans Produce the Final Artifact.}
Figure~\ref{fig:richer-settings-curves}(a) shows the results in co-creation settings with an agent using a single policy. \condAIDivHumanConvOutLow{}, where the agent provides ideas and humans produce the final artifacts, exceeds \condHumanOnly{} diversity in \domainHypothesisGeneration{} and falls slightly below it in \domainStoryWriting{} at full adoption. \condHumanDivAIConvOutLow{}, where humans provide ideas and the agent produces the final artifacts, achieves lower diversity than \condAIDivHumanConvOutLow{} in both domains. This finding indicates that settings where humans produce the final artifact generally preserve more diversity than those where the agent does.

\textbf{Mixture-based Agent Design Broadly Mitigates Diversity Collapse.} Figure~\ref{fig:richer-settings-curves}(b) shows the results in co-creation settings with an agent using a mixture-based policy. We found that co-creating with this agent achieves higher diversity in most settings, compared to co-creating with the agent using a single policy in Figure~\ref{fig:richer-settings-curves}(a). Notably, several curves peak at intermediate adoption rates, suggesting that partial adoption can yield greater diversity than full adoption. These results suggest that a more diverse agent policy can help mitigate diversity loss, but the resulting diversity--utility tradeoffs depend on the interaction mode, domain, and adoption level.

\section{RQ3: Predicting the Dynamics of Collective Diversity}
\label{sec:evaluation3}

\subsection{Formulations for Learning Predictive Functions}\label{subsec:evaluation3-formulation}

\looseness-1\textbf{Prediction Goal.} Given a task $\task$ and a collection of artifacts from human users acting alone without AI assistance $\mathcal{O}^T_{\mathcal{H}, \text{none}}$, we want to understand how different deployment configurations $d=(\agent, x)$ would affect the collective diversity when users switch to co-create with the agent $\agent$ at rate $x$. Given some historical data about co-creation under other settings with various agent designs, we seek to learn such predictive functions that can provide insights into this for the current task $\task$ before deploying the target configuration. These
predictors allow us to ask how diversity would scale as adoption increases, and whether different agent designs or interaction modes are likely to preserve or reduce collective diversity, similar in spirit to the scaling laws for generative
models~\cite{DBLP:journals/corr/abs-2001-08361,10.5555/3600270.3602446,DBLP:conf/iclr/WuSLWY25,qin2025scaling}.

\looseness-1\textbf{Basic Predictive Functional Form.} Guided by Eq.~\ref{eq:diversity-adoption-quadratic} and as noted in Section~\ref{subsec:evaluation1-formulation}, the collective diversity is a quadratic function of the adoption rate. However, it depends on the co-creation setting in a non-trivial way through the distribution over artifacts $\mathcal{O}^T_{\mathcal{H}, \agent}$. Without any additional assumptions about the agent design, we consider a basic quadratic predictor $\funcOneParam(\adoptionRate) = a + bx + cx^2$. At zero adoption, all artifacts are human-only artifacts, so $a = D_H := \metricDiversity(\mathcal{O}^T_{\mathcal{H}, \text{none}})$. We learn the remaining parameters $b$ and $c$ from historical data, capturing the overall trend over adoption rates for a new configuration.

\looseness-1\textbf{Richer Predictive Functional Forms.} While $\funcOneParam(\adoptionRate)$ provides a simple form, it does not allow us to capture any properties of the agent $A$ for which we seek to make predictions. To this end, we will develop functional forms that can model richer co-creation settings and can be learned from data. Below we provide a high-level overview of obtaining two new functional forms; see Appendix~\ref{app:prediction} for further details.
We first make a simple modeling choice for $\mathcal{O}^T_{\mathcal{H}, \agent}$ as follows.
We model the distribution $P(O^T_{\mathcal{H}, \agent})$ by decomposing it into two distribution sources of human-only artifacts $P(O^T_{\mathcal{H}, \text{none}})$ and agent-only artifacts $P(O^T_{\text{none}, \agent})$. More concretely, let $\lambda \in [0,1]$ represent the agent's influence on the co-created artifact distribution, with larger values of $\lambda$ indicating that co-created artifacts move further from the human-only distribution toward the agent distribution. This in turn gives us a mixture with weights $(1-\lambda)$ for $P(O^T_{\mathcal{H}, \text{none}})$ and $\lambda$ for $P(O^T_{\text{none}, \agent})$. With this mixture, the $\metricDiversity(A, x;T)$ from Eq.~\ref{eq:diversity-adoption-quadratic} can now be expressed through the following agent design factors, namely $\metricDiversity(\mathcal{O}^T_{\text{none}, \agent})$,  $\text{Dis}(\mathcal{O}^T_{\mathcal{H}, \text{none}}, \mathcal{O}^T_{\text{none}, \agent})$, and $\lambda$. With a short notation of $D_A :=  \metricDiversity(\mathcal{O}^T_{\text{none}, \agent})$ and $D_{AH} :=  \text{Dis}(\mathcal{O}^T_{\mathcal{H}, \text{none}}, \mathcal{O}^T_{\text{none}, \agent})$, this brings us to two richer functional forms that carry information about the agent design as stated below: $\funcTwoParam(\adoptionRate, D_A, D_{AH})$ that does not explicitly use $\lambda$, and $\funcThreeParam(\adoptionRate, D_A, D_{AH}, \lambda)$ that also uses $\lambda$. In the following, we state these two functional forms along with $\funcOneParam(\adoptionRate)$ introduced earlier:

\vspace{-6mm}
\begin{align}
\funcOneParam(\adoptionRate) &= D_H + b{x} + c{x}^2, \label{eq:f1-form} \\
\funcTwoParam(\adoptionRate, D_A, D_{AH}) &= D_H + b(D_{AH}-D_H)x + c(D_H+D_A-2D_{AH})x^2, \label{eq:f2-form} \\
\funcThreeParam(\adoptionRate, D_A, D_{AH}, \lambda) &= D_H + b\lambda(D_{AH}-D_H)x + c\lambda^2(D_H+D_A-2D_{AH})x^2. \label{eq:f3-form}
\end{align}

Overall, the new functional forms move from a predictor that uses only the adoption rate to predictors informed by agent design. When fitting the parameters $b$ and $c$ from historical data, we pre-compute and fix any terms associated with the agent, including $D_A$, $D_{AH}$, and ${\lambda}$.


\subsection{Evaluation Setup}
\label{subsec:evaluation3-setup}
Next, we describe our evaluation setup, including model fitting, prediction, and evaluation procedures. In the following, we distinguish the target agent for which we aim to predict collective diversity from the other agents for which historical co-creation data are available.

\looseness-1\textbf{Data for fitting.}
We fit the predictive models using historical data from the other agent settings in the same domain, without any data for the target agent. For each historical setting, we compute the corresponding source quantities $D_H,D_A,D_{AH}$, agent influence $\widehat{\lambda}$, and construct adoption--diversity pairs from the sampled artifact sets for evenly spaced adoption rates $x\in\{0,0.01,\ldots,1\}$. Each setting has its own agent influence estimate $\widehat{\lambda}$, learned so that the human--agent mixture best matches the distribution of artifacts from co-creation  (see Appendix~\ref{app:prediction} for details). For fitting, $\funcOneParam$ uses human-only diversity $D_H$ to learn a general adoption--diversity trend from these pairs. $\funcTwoParam$ additionally uses agent-only diversity $D_A$ and human--agent dissimilarity $D_{AH}$, allowing the predicted trend to depend on the agent policy. $\funcThreeParam$ further uses estimated agent influence $\widehat{\lambda}$, allowing the prediction to account for the interaction mode. 
We fit each predictive model's coefficients $b,c$ using observed collective diversity by ridge regression~\cite{murphy2012machine}. Each model's coefficients are shared across settings and tasks within a domain.


\looseness-1\textbf{Data for prediction.}
Next we discuss how to do predictions for a target agent using the fitted coefficients $b,c$, along with some relevant quantities for the target agent. More concretely, apart from using the fitted coefficients, each predictive model uses the following data as input:

\begin{itemize}[leftmargin=*]
    \item $\funcOneParam$ uses only the adoption rate $x$ and the corresponding human-only diversity $D_H$. 
    \item $\funcTwoParam$ additionally uses two quantities $D_A,D_{AH}$ that are associated with the target agent. 
    \item $\funcThreeParam$ additionally uses $\widehat{\lambda}$ transferred from another setting which has the same interaction mode but a different agent policy compared to the target agent. For instance, when \condHumanDivAIConvOutHigh{} is the target, we can use $\widehat{\lambda}$ estimated from \condHumanDivAIConvOutLow{}. This transfer assumes that influence learned under one policy is informative about the same interaction mode under another policy.
\end{itemize}

\looseness-1\textbf{Evaluation.}
Within each domain, we conduct leave-one-out evaluation where we exclude one setting from fitting across all four tasks. We fit the predictive models using data from the remaining nine settings and predict collective diversity for the excluded setting. The combinations of five interaction modes and two agent policies give ten evaluation folds per domain. For each fold, we compute the Root Mean Squared Errors (RMSEs) over adoption rates separately for each task, then average the RMSEs for the four tasks. We report the mean and standard error of these averages across folds.

\subsection{Results}
\label{subsec:evaluation3-results}

\textbf{Predictive Accuracy.}
Table~\ref{tab:prediction-rmse} compares prediction error across held-out settings in both domains. Incorporating agent diversity $D_A$ and human--agent dissimilarity $D_{AH}$ in $\funcTwoParam$ reduces error relative to the adoption-only predictor $\funcOneParam$. Adding estimated agent influence in $\funcThreeParam$ further improves average accuracy, yielding the lowest mean error in both domains. These results suggest that predicting collective diversity benefits from accounting for both the agent's output distribution and its influence on co-created artifacts.

\begin{figure}[]
    \centering
    \includegraphics[width=\textwidth]{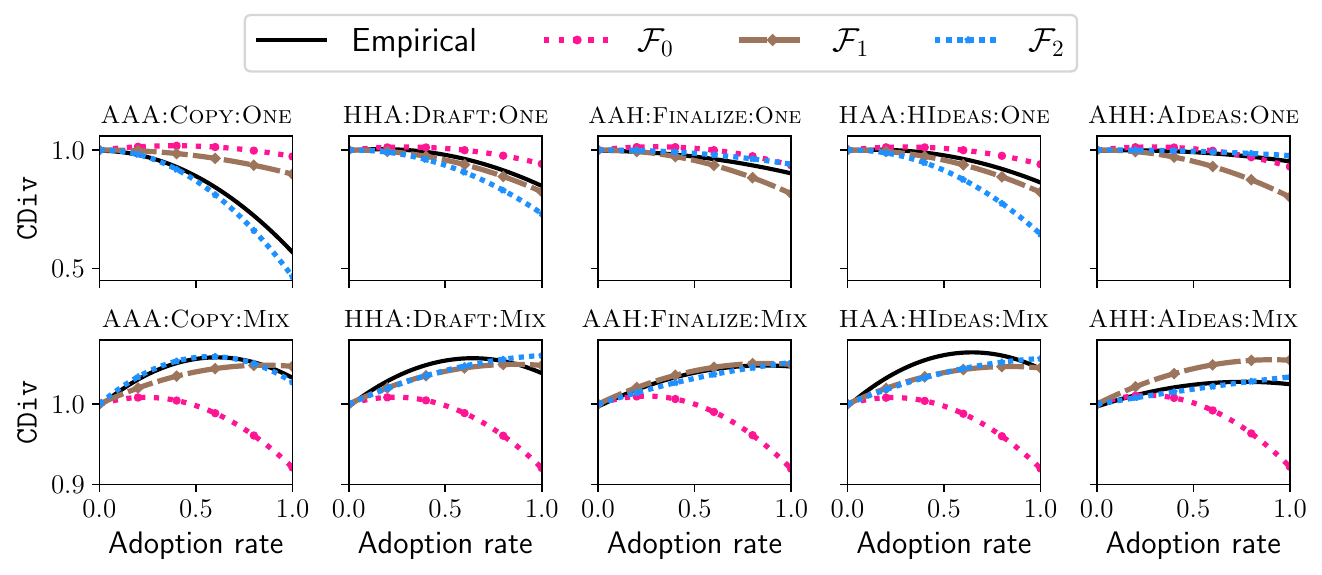}
    \caption{\looseness-1 (RQ3) Empirical and predicted curves for held-out conditions in \domainStoryWriting{}.}
    \label{fig:prediction-curves}
\end{figure}

\textbf{Predicted Curves.} Figure~\ref{fig:prediction-curves} qualitatively compares predicted and empirical diversity curves for \domainStoryWriting{}.
$\funcOneParam$ predicts a similar overall decline across settings, failing to distinguish
differences between agent policies. Both richer predictors capture the contrasting trends where
overall diversity reduces when co-creating with an agent using a single policy and maintains when co-creating with an agent using a mixture-based policy.


\begin{table}[]
    \centering
    \caption{(RQ3) Predictive performance on held-out settings. Each fold holds out one co-creation setting for evaluation. We average task-level root mean squared errors (RMSEs) within each fold and report the mean across folds, with the standard error (SE).}
    \label{tab:prediction-rmse}
    \renewcommand{\arraystretch}{1.4}
\centering
\begin{tabular}{cccc}
\toprule
\shortstack[l]{\textbf{Predictive Model}} & \textbf{\domainHypothesisGeneration} & \textbf{\domainStoryWriting} & \textbf{Average}\\
\midrule
$\mathcal{F}_{0}$ & 0.029 (0.007) & 0.034 (0.008) & 0.031 \\
$\mathcal{F}_{1}$ & 0.023 (0.008) & 0.020 (0.007) & 0.022 \\
\textbf{$\mathcal{F}_{2}$} & \textbf{0.015 (0.002)} & \textbf{0.016 (0.005)} & \textbf{0.016} \\
\bottomrule
\end{tabular}

\end{table}

\section{Concluding Discussions} \label{sec:conclusion}
We studied how collective diversity changes as AI adoption increases across a population. We found that AI agents can improve artifact quality and reduce user effort at the cost of declining collective diversity. We showed that more fine-grained interaction modes and more diverse agent policies may mitigate this decline. Furthermore, we derived predictive models that can estimate collective diversity across different levels of AI adoption, informing the design of agents that preserve collective diversity.

\looseness-1Next, we discuss a few limitations of our work and outline a future plan to address them.
First, the examined co-creation settings are based on structured interfaces with distinct stages, whereas human-AI co-creation may unfold through less structured, conversational interactions; future work could examine whether our findings generalize to conversational co-creation settings.
Second, the adoption rate in our study is simulated by sampling and mixing artifacts collected under controlled co-creation settings, rather than observing adoption where users freely choose whether to adopt AI agents; future work could examine deployment settings in which adoption rates are measured directly from users' actual adoption of AI agents (see Appendix~\ref{app:endogenous-adoption}).
Third, our predictive functions are fitted and evaluated within the domains and co-creation settings in this study, and may not generalize to new settings; future work could develop richer predictors that incorporate additional features about the domain or co-creation setting.

As broader implications of our work, we do not anticipate any direct negative societal impacts arising from this research -- our findings can inform the design of AI agents that help preserve collective diversity in human-AI co-creation when these agents are deployed at scale.

\clearpage
\begin{ack}
Manh Hung Nguyen acknowledges support for a research visit to the University of Vienna from March to May 2026 through ELSA – European Lighthouse on Secure and Safe AI, funded by the European Union under Grant Agreement No. 101070617.
\end{ack}

\bibliographystyle{unsrtnat}
\bibliography{main}

\clearpage
\appendix

\allowdisplaybreaks


\section{Table of Contents}
\label{app:toc}

In this section, we briefly describe the content provided in the paper's appendices.

\begin{itemize}
    \item Section~\ref{app:domains} provides additional details about the creative domains and study statistics.
    \item Section~\ref{app:prediction} provides additional details about the richer predictive functional forms.
    \item Section~\ref{app:additional-experiments} provides additional studies and experiments.
    \item Section~\ref{app:instructions} provides additional details about the interfaces and instructions of our studies.
    \item Section~\ref{app:prompts} provides additional details about the prompts used to generate agent responses.
\end{itemize}

\clearpage

\section{Additional Details about Domains and Study Statistics}
\label{app:domains}

We provide the tasks and example answers from \condHumanDivAIConvOutHigh{} for the two creative domains, \domainHypothesisGeneration{} and \domainStoryWriting{}, in Table~\ref{tab:domains}.


\begin{table}[ht!]
    \centering
    \footnotesize
    \setlength{\tabcolsep}{4pt}
    \renewcommand{\arraystretch}{1.1}

    \caption{Tasks and example co-created answers in the two creative domains.}

    \begin{subtable}[t]{\textwidth}
        \centering
        \caption{\domainHypothesisGeneration{}}
        \begin{tabular}{p{0.26\linewidth}p{0.66\linewidth}}
        \toprule
        \textbf{Task} & \textbf{Example answer} \\
        \midrule
        You just started going to a new school. You notice that more students sing in the hallways at this school than at your old school. What hypotheses do you have about why that is? &
        The school was built on the former site of a music academy whose acoustically perfect hallway design unconsciously triggers students' desire to vocalize. \\
        \midrule
        You notice that dogs seem to like one of your friends, but cats seem to like another friend. What hypotheses do you have about why that is? &
        Each friend unconsciously mimics the body language of the species they were raised around as children, making dogs and cats recognize them as `familiar kin' rather than strangers. \\
        \midrule
        You notice that the water in one lake is warmer than the water in another lake even though they both get the same amount of sunlight. What hypotheses do you have about why that is? &
        The warmer lake contains a higher concentration of dissolved organic compounds from decaying vegetation that absorb and trap infrared radiation like a submerged greenhouse, re-radiating heat back into the water column. \\
        \midrule
        You look outside one night and see stars disappearing one by one from the sky. What hypotheses do you have about why that is? &
        The stars are disappearing because Earth has unexpectedly entered a region of localized spacetime curvature~-- a `gravitational wake' left by a primordial black hole passing through our solar system, causing distant starlight to lens away from our line of sight in a rolling wave pattern. \\
        \bottomrule
        \end{tabular}
        
        \label{tab:domains-hg}
    \end{subtable}

    \vspace{0.25cm}

    \begin{subtable}[t]{\textwidth}
        \centering
        \caption{\domainStoryWriting{}}
        \begin{tabular}{p{0.26\linewidth}p{0.66\linewidth}}
        \toprule
        \textbf{Task} & \textbf{Example answer} \\
        \midrule
        Write a 3--5 sentence story using: stamp, letter, send. &
        The last stamp on earth was tucked in a tin box beneath old Mr. Greeley's mattress. He'd waited 60 years for the right moment to send a letter to his son, gone since the war. When he finally placed the stamp~-- a faded blue thing with golden edges~-- on the envelope, the postmaster refused to take it. `That one doesn't exist anymore,' he said. That night, Mr. Greeley lit a candle, dropped the letter into the flame, and watched it vanish~-- written on the smoke was a single word: `Received.' \\
        \midrule
        Write a 3--5 sentence story using: petrol, diesel, pump. &
        My students were fascinated when I brought an AI robot into our science lesson about energy sources. The robot happily announced it ran on neither petrol nor diesel, making it the cleanest visitor our classroom had ever seen. One curious first-grader asked if robots needed a pump to refuel. The AI giggled and explained that its fuel was simply knowledge shared by children every day. \\
        \midrule
        Write a 3--5 sentence story using: gloom, payment, exist. &
        The office was steeped in gloom, not from lack of light, but because no one had received payment in weeks. A junior developer whispered to another, `Do you think this place even exists? I can't find it on any map.' Suddenly, the lights flickered, and the nameplate on the door changed: `Payment Processing~-- Now and Forever.' We realized we were ghosts, bound by unpaid wages, forced to exist in the system until our debt was cleared. \\
        \midrule
        Write a 3--5 sentence story using: organ, empire, comply. &
        The emperor knelt before the great organ in the cathedral, its pipes towering like pillars of his crumbling empire. `Play,' he commanded, but the organ replied in a low hum, `I will not comply unless the people's songs are heard.' For the first time, the ruler trembled~-- not from rage, but fear that his empire would fall to silence. The next morning, the organ played a new anthem, one the people had sung in secret for years. \\
        \bottomrule
        \end{tabular}
        
        \label{tab:domains-csw}
    \end{subtable}

    \label{tab:domains}
\end{table}

Table~\ref{tab:study-statistics} shows the statistics of the co-creation studies. Table~\ref{tab:annotation-statistics} shows the statistics of the quality annotation studies.

\begin{table}[!htbp]
    \centering
    \caption{Statistics of the co-creation studies.}
    \label{tab:study-statistics}
    \resizebox{0.99\linewidth}{!}{%
    \begin{tabular}{@{}lrr@{}}
        \toprule
        & \domainHypothesisGeneration{} & \domainStoryWriting{} \\
        \midrule
        Tasks & 4 & 4 \\
        Co-creation settings & 11 & 11 \\
        Artifacts per task and setting\footnotemark & 50 & 50 \\
        Total artifacts & 2,200 & 2,200 \\
        \midrule
        Co-creation settings with participants & 9 & 9 \\
        Total artifacts in settings with participants & 1,800 & 1,800 \\
        \midrule        
        Tasks assigned to each participant & 4 & 1 \\
        Allocated time per participant (minutes) & 14 & 10 \\
        Total number of unique participants & 450 & 1,800 \\
        \bottomrule
    \end{tabular}
    }
\end{table}
\footnotetext{The number of artifacts varies slightly across tasks and settings, and 50 is the average.}

\begin{table}[!htbp]
    \centering
    \caption{Statistics of the quality annotation studies.}
    \label{tab:annotation-statistics}
    \resizebox{0.99\linewidth}{!}{%
    \begin{tabular}{@{}lrr@{}}
        \toprule
        & \domainHypothesisGeneration{} & \domainStoryWriting{} \\
        \midrule
        Artifacts  & 2,200 & 2,200 \\
        Annotations per artifact\footnotemark & 1 & 1 \\
        Total annotations & 2,220 & 2,214 \\
        \midrule
        Artifacts assigned to each annotator \qquad \qquad & 12 & 3 \\
        Time per annotator (minutes) & 12 & 8 \\
        Total number of unique annotators & 185 & 738 \\
        \bottomrule
    \end{tabular}
    }
\end{table}
\footnotetext{A few artifacts received two annotations, and we average their quality scores. Consequently, the ``Total annotations'' is more than 2200 in this table.}

\clearpage

\section{Additional Details on Predictive Functional Forms}
\label{app:prediction}
\subsection{Obtaining Richer Functional Forms}
\label{app:prediction-richer-forms}

Here we provide some further details on how to obtain the richer functional forms, expanding on the discussion from Section~\ref{subsec:evaluation3-formulation}. 
To arrive at functional forms that can model richer co-creation settings and can be learned from data, we will first make a simple modeling choice for $\mathcal{O}^T_{\mathcal{H}, \agent}$.
We model the distribution $P(O^T_{\mathcal{H}, \agent})$ by decomposing it into three distribution sources: human-generated artifacts $P(O^T_{\mathcal{H}, \text{none}})$, agent-generated artifacts $P(O^T_{\text{none}, \agent})$, and interaction-driven artifacts that are not well explained by a simple human-agent mixture. More concretely, let $\lambda \in [0,1]$ represent the agent's influence on the co-created artifact distribution with larger values of $\lambda$ indicating that co-created artifacts move further from the human-only distribution toward the agent distribution; $\rho \in [0,1]$ captures the share of co-created artifacts generated from a novel distribution of artifacts that emerge from the interaction, and not captured by a linear mixture of human and agent distributions. This in turn gives us $P(O^T_{\mathcal{H}, \agent})$ as a mixture with the following weights: $(1-\rho) (1-\lambda)$ for $P(O^T_{\mathcal{H}, \text{none}})$, $(1-\rho) \lambda$ for $P(O^T_{\text{none}, \agent})$, and $\rho$ for $P(O^T_{\text{none}, \text{none}})$ for novel artifacts.

When considering $\metricDiversity(A, x;T)$ from Eq.~\ref{eq:diversity-adoption-quadratic}, we have $(1-x)$ fraction of artifacts from $P(O^T_{\mathcal{H}, \text{none}})$ and $x$ fraction of artifacts from $P(O^T_{\mathcal{H}, \agent})$. To put it together, we can consider this as a mixture with the following weights:
\begin{itemize}
    \item $P(O^T_{\mathcal{H}, \text{none}})$ with $(1-x) + (1-\rho) (1-\lambda) x$ 
    \item  $P(O^T_{\text{none}, \agent})$ with $(1-\rho) \lambda x$
    \item $P(O^T_{\text{none}, \text{none}})$ with $\rho x$ 
\end{itemize}

We plug this mixture distribution of sampling artifacts into $\metricDiversity(A, x;T)$ from Eq.~\ref{eq:diversity-adoption-quadratic}. Next, to integrate the effects of $P(O^T_{\text{none}, \text{none}})$ which is unknown, we will simply set it to a uniform distribution over the artifacts space.  Finally, we also absorb the mixing factor of $\rho$ directly into the parameters, which provides us a functional form $\mathcal{F}$ with $7$ parameters along with $\lambda$ as stated below:

\begin{align*}   
\mathcal{F}(\adoptionRate, D_A, D_{AH}, \lambda) =  \ D_H &+ \big(b_1 - b_2D_H  +  
    b_3\lambda(D_{AH} - D_H)\big){x}   \notag\\
    & +  \big(c_1 + 
    c_2D_H + c_3\lambda(D_H - D_{AH}) + c_4\lambda^2(D_H+ D_A - 2D_{AH})\big){x}^2,
\end{align*}

where we used a short notation of $D_H :=  \metricDiversity(\mathcal{O}^T_{\mathcal{H}, \text{none}})$, $D_A :=  \metricDiversity(\mathcal{O}^T_{\text{none}, \agent})$, and $D_{AH} :=  \text{Dis}(\mathcal{O}^T_{\mathcal{H}, \text{none}}, \mathcal{O}^T_{\text{none}, \agent})$. We collapse several of these parameters to obtain a form $\funcThreeParam$ with $2$ parameters along with $\lambda$ (see Eq.~\ref{eq:f3-form}), to make it more directly comparable to $\funcOneParam$ in terms of predictive power which has $2$ parameters (see Eq.~\ref{eq:f1-form}). When we further absorb $\lambda$ into the parameters, we obtain a simpler functional form $\funcTwoParam$ with  $2$ parameters (see Eq.~\ref{eq:f2-form}).

These different functional forms carry information about the agent design, in contrast to the basic $\funcOneParam$. While additional information is potentially useful, there is also a trade-off as these richer forms require estimating the agent-specific information from data. We will look at the predictive power of these functional forms in Section~\ref{subsec:evaluation3-results}. 

\subsection{Estimating Agent Influence}
\label{app:prediction-influence}

Next we discuss how to estimate agent influence and also illustrate the estimated values for different co-creation settings. We estimate one influence $\widehat{\lambda}$ for each co-creation setting, jointly across its four tasks. In addition to $D_H$, $D_A$, and $D_{AH}$ introduced earlier, for brevity we introduce some additional short notation as follows: $D_{C} :=  \metricDiversity(\mathcal{O}^T_{\mathcal{H}, \agent})$,  $D_{CH} :=  \text{Dis}(\mathcal{O}^T_{\mathcal{H}, \text{none}}, \mathcal{O}^T_{\mathcal{H}, \agent})$, and $D_{CA} :=  \text{Dis}(\mathcal{O}^T_{\text{none}, \agent}, \mathcal{O}^T_{\mathcal{H}, \agent})$. 
$D_C$ denote the diversity of the setting's co-created artifacts, and $D_{CH}$ and $D_{CA}$ their dissimilarities to human-only and agent-only artifacts, respectively.
For a candidate influence $\lambda$, the two-source human--agent mixture predicts these quantities as follows:
\begin{align*}
\widehat{D}_C(\lambda)&=(1-\lambda)^2D_H+\lambda^2D_A+2\lambda(1-\lambda)D_{AH},\\
\widehat{D}_{CH}(\lambda)&=(1-\lambda)D_H+\lambda D_{AH},\\
\widehat{D}_{CA}(\lambda)&=(1-\lambda)D_{AH}+\lambda D_A.
\end{align*}
\looseness-1We estimate $\widehat{\lambda}$ by grid search over $\lambda\in\{0,0.001,\ldots,1\}$, minimizing the summed squared relative error between predicted and observed values across all three quantities and all four tasks.
Once we obtain $\widehat{\lambda}$, then we use it to further estimate $\widehat{\rho}$ using a similar procedure that minimizes error between predicted and observed values across relevant quantities connecting $\lambda$ and $\rho$ introduced in Section~\ref{app:prediction-richer-forms}.

\begin{figure}[htb]
    \centering
    \includegraphics[width=0.7\linewidth]{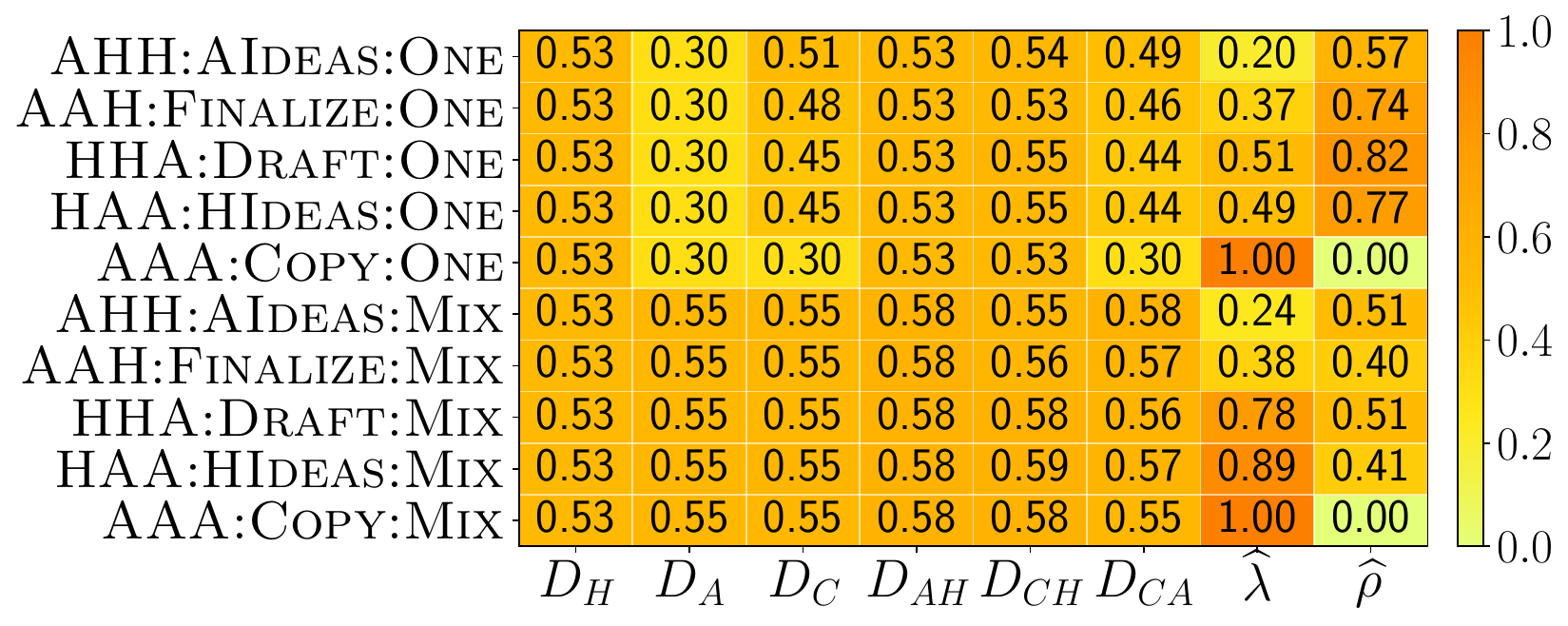}
    \caption{Diversity quantities, agent influence $\widehat{\lambda}$, and novelty share $\widehat{\rho}$ in \domainStoryWriting{}.}
    \label{fig:influence-heatmap}
\end{figure}

We summarize the estimated quantities for each co-creation setting in Fig.~\ref{fig:influence-heatmap}. Agent influence is higher when the agent controls the final output. Compared with policy \textsc{One}, policy \textsc{Mix} has higher agent influence $\lambda$ but lower estimated $\rho$ across all four interaction modes.

\subsection{Agent Design and Collective Diversity}
\label{app:prediction-contours}

In the experiments conducted in Section~\ref{sec:evaluation3}, we considered leave-one-out evaluation where we exclude one setting
from fitting, fit the predictive models using data from the remaining settings, and predict collective diversity for the excluded setting. Here, we conduct additional experiments where we predict collective diversity for new agent designs by varying the properties $D_A$ (agent-only diversity), $D_{AH}$ (human--agent dissimilarity), and $\lambda$ (agent's influence).

\begin{figure}[h!]
    \centering
    \includegraphics[width=0.8\textwidth]{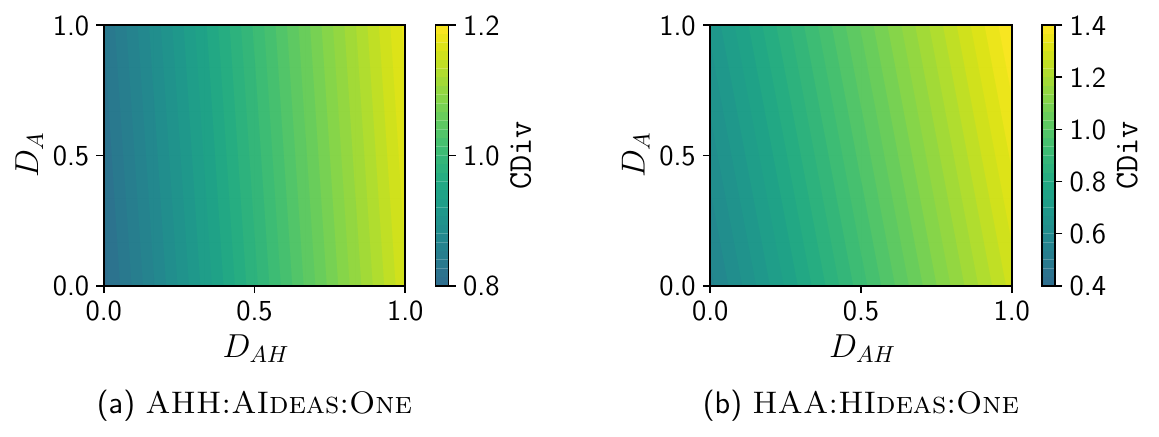}
    \caption{\looseness-1Collective diversity predicted by $\funcThreeParam$ for {\condAIDivHumanConvOutLow} (a) and {\condHumanDivAIConvOutLow} (b) in \domainStoryWriting{} as agent diversity $D_A$ and human--agent dissimilarity $D_{AH}$ vary. Predictions fix the adoption rate at $x=0.5$ and use each setting's estimated agent influence $\widehat{\lambda}$. Lighter colors indicate higher collective diversity, normalized by \condHumanOnly{}.}
    \label{fig:contours-diversity-dissimilarity}
\end{figure}

\looseness-1\textbf{Agent Diversity and Human--Agent Dissimilarity.} We use the fitted $\funcThreeParam$ (Eq.~\ref{eq:f3-form}) to compare how collective diversity responds to agent diversity $D_A$ and human--agent dissimilarity $D_{AH}$. More concretely, for {\condAIDivHumanConvOutLow} and {\condHumanDivAIConvOutLow} in \domainStoryWriting{}, we fix the adoption rate at $\adoptionRate=0.5$ and agent influence at each setting's estimated $\widehat{\lambda}$ (Section~\ref{app:prediction-influence}). Then, we vary $D_A$ and $D_{AH}$  as illustrated in Figure~\ref{fig:contours-diversity-dissimilarity}. In both settings, collective diversity is more sensitive to human--agent dissimilarity than to agent diversity. 

\begin{figure}[h!]
    \centering
    \includegraphics[width=0.8\textwidth]{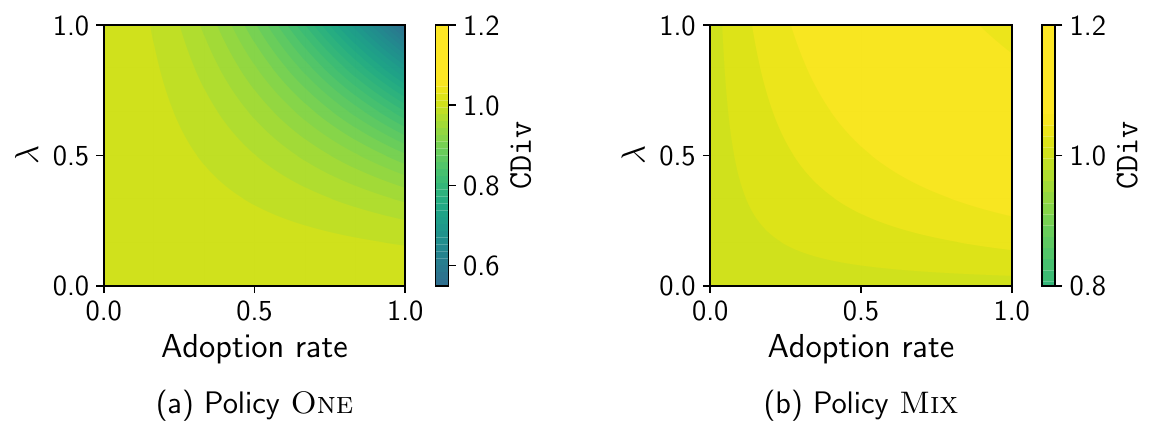}
    \caption{\looseness-1Collective diversity predicted by $\funcThreeParam$ in \domainStoryWriting{} as adoption rate $x$ and agent influence $\lambda$ vary. 
    $D_H$, $D_A$, and $D_{AH}$ values are fixed, based on \textsc{One} and \textsc{Mix} policy settings. Lighter colors indicate higher collective diversity, normalized by \condHumanOnly{}.}
    \label{fig:contours-adoption-influence}
\end{figure}

\textbf{Adoption Rate and Agent Influence.} For a given agent policy, collective diversity depends on how many users adopt the agent and on how strongly the agent shapes each co-created artifact. 
Intuitively, an agent that many users adopt but only lightly shapes their artifacts may change collective diversity as much as an agent that few users adopt but strongly shapes their artifacts. 
More concretely, for \textsc{One} and \textsc{Mix} policy settings in \domainStoryWriting{}, we fix $D_H$, $D_A$, and $D_{AH}$ to their respective values in these settings. Then, we vary agent influence $\lambda$ and adoption rate $\adoptionRate$ as illustrated in Figure~\ref{fig:contours-adoption-influence}. 
When considering \textsc{One} policy (low agent-only diversity $D_A$), increasing adoption with strong agent influence showcases the trend of reduced collective diversity. When considering \textsc{Mix} policy (high agent-only diversity $D_A$), increasing adoption largely preserves collective diversity, even under strong agent influence.

\clearpage
\section{Additional Experiments}
\label{app:additional-experiments}

\subsection{Endogenous Adoption}
\label{app:endogenous-adoption}

Our main experiments vary adoption by mixing artifacts from fixed human-only conditions and conditions with AI. In practice, users choose whether to use AI, and those who accept assistance may differ from those who decline it. We therefore test whether outcomes from mixing-based adoption agree with those from endogenous adoption where participants can make this choice. We ran an additional study on the four tasks in \domainHypothesisGeneration{} with the single-model policy \textsc{One}. This specific study used GPT-5.4~\citep{openai2026gpt54} as the underlying model.

\paragraph{Study design: Endogenous adoption.} We used two settings, \condAIOnlyLow{} and \condAIDivConHumanOutLow{} as in the main studies (cf.\ Section~\ref{sec:evaluation1} and Section~\ref{sec:evaluation2}). We also used \condAIDivHumanConvOutLow{} where AI agent supplies ideas once, and unlike in the main studies, participants cannot request new ideas. The participant then writes the final answer. Participants were assigned to conditions corresponding to these settings and could decide whether to use the offered AI assistance. We recruited 50 new participants on Prolific for each of these conditions. After seeing a brief task description, participants chose whether to accept the offered assistance or work without the agent. Each participant then completed the four tasks. The observed adoption rate is the fraction of participants who accepted assistance.

\paragraph{Study design: Mixing-based adoption.} We separately recruited participants and assigned them to \condHumanOnly{}, \condAIDivHumanConvOutLow{} (with ideas supplied once), and \condAIDivConHumanOutLow{}. Participants in the conditions with AI agent were asked to use the offered AI assistance without an option to opt out. For \condAIOnlyLow{}, we generated and used the artifacts directly from the agent without recruiting participants. 

\begin{table}[!htbp]
    \centering
    \setlength{\tabcolsep}{4pt}
    \caption{Endogenous adoption vs.\ mixing-based adoption in \domainHypothesisGeneration{} under the single-model policy \textsc{One}.
    Obs.\ denotes results when participants decide whether to use AI; Mix.\ denotes results from mixing.
    All values are normalized by \condHumanOnly{}.}
    \label{tab:endogenous-adoption}
    \begin{tabular}{@{}lc*{6}{r}@{}}
        \toprule
        & & \multicolumn{2}{c}{$\metricDiversity$}
          & \multicolumn{2}{c}{$\metricQuality$}
          & \multicolumn{2}{c}{$\metricEffort$} \\
        \cmidrule(lr){3-4}\cmidrule(lr){5-6}\cmidrule(lr){7-8}
        Setting & Observed Adoption Rate & Obs. & Mix. & Obs. & Mix. & Obs. & Mix. \\
        \midrule
        \condAIOnlyLow{}           & 0.50 & 0.84 & 0.83 & 1.24 & 1.31 & 0.60 & 0.50 \\
        \condAIDivConHumanOutLow{}  & 0.53 & 0.94 & 0.91 & 1.03 & 1.11 & 0.79 & 0.86 \\
        \condAIDivHumanConvOutLow{} & 0.67 & 1.05 & 1.04 & 0.95 & 1.05 & 0.88 & 0.78 \\
        \bottomrule
    \end{tabular}
\end{table}

\paragraph{Results.} We compared the observed results in endogenous adoption with results from mixing-based adoption, following the evaluation procedure in Sections~\ref{sec:evaluation1} and~\ref{sec:evaluation2}. Table~\ref{tab:endogenous-adoption} shows that, given the observed adoption rates, collective diversity in mixing-based adoption closely mirrors collective diversity in endogenous adoption.
Observed quality is lower than the values from mixing in all three settings. While the rankings between settings in terms of quality and effort largely remain the same, there are some differences between observed quality and effort values compared to mixing-based values.
This suggests that, in practice, participants who accept AI-assistance may differ from those who decline, which changes who contributes to each artifact pool, and allowing a choice may also change how participants work. Overall, the diversity results support the use of mixing-based adoption used in our main studies.
\subsection{Comparison of Diversity Measures with Human Evaluation}
\label{app:human-rated-diversity}

We test whether $\metricDiversity$ agrees with human judgments of diversity. We also compare it with the Vendi score~\citep{DBLP:journals/tmlr/FriedmanD23} and a lexical diversity measure. We ran an additional study on \domainHypothesisGeneration{}, separate from the main studies and from the study in Appendix~\ref{app:endogenous-adoption}, and used two of its tasks. The single-model policy \textsc{One} uses GPT-5.4~\citep{openai2026gpt54} as the underlying model without personas, and the mixture-based policy \textsc{Mix} samples personas and ten open-weight models, namely \texttt{Qwen3.5-27B}, \texttt{Qwen3.5-122B-A10B}~\citep{qwen3.5}, \texttt{gemma-3-12b-it}~\citep{DBLP:journals/corr/abs-2503-19786}, \texttt{gemma-4-E4B-it}~\citep{DBLP:journals/corr/abs-2607-02770}, \texttt{Mistral-Small-3.1-24B-Instruct-2503}~\citep{mistral-small-3-1}, \texttt{Mistral-Large-3-675B-Instruct-2512}~\citep{mistral-3}, \texttt{Kimi-K2-Thinking}~\citep{kimi-k2-thinking}, \texttt{Kimi-K2.5}~\citep{DBLP:journals/corr/abs-2602-02276}, \texttt{GLM-4.6}~\citep{DBLP:journals/corr/abs-2508-06471}, and \texttt{GLM-5.1}~\citep{glm-5.1}.

\paragraph{Study design: Artifact collection.} We used three settings, \condHumanOnly{}, \condAIOnlyLow{}, and \condAIOnlyHigh{}, as in the main studies (cf.\ Section~\ref{sec:evaluation1} and Section~\ref{sec:evaluation2}). We also used \condAIDivHumanConvOutLow{} and \condAIDivHumanConvOutHigh{} where AI agent supplies ideas once, and unlike in the main studies, participants cannot request new ideas. The participant then writes the final answer. We recruited participants on Prolific and assigned each of them to one condition. For \condAIOnlyLow{} and \condAIOnlyHigh{}, we generated and used the artifacts directly from the agent without recruiting participants. This gives five conditions per task.

\paragraph{Study design: Human clustering.} We sampled 20 artifacts without replacement from each task and condition, which gave 10 batches (2 tasks $\times$ 5 conditions) and 200 artifacts in total. We then collected new human annotations on these batches. Three annotators independently clustered every batch, grouping answers that expressed the same or a very similar hypothesis. They saw the task prompt and answers, but not the source condition, and could create as many clusters as needed. We measured diversity from human judgments as the mean number of occupied clusters across the three annotators. The mean pairwise Pearson correlation between annotators' raw cluster counts across the 10 batches was $r=0.81$.

\begin{table}[h!]
    \centering
    \setlength{\tabcolsep}{4pt}
    \caption{Diversity metrics comparison.
    Raw values are means across the two tasks.
    For normalized values (Norm.), we first divide each task's score by its human-only baseline, then average across tasks.
    Human scores also average across the three annotators.}
    \label{tab:human-rated-diversity}
    \begin{tabular}{@{}l r rr rr rr@{}}
        \toprule
        & $\metricDiversity$ & \multicolumn{2}{c}{Human}
          & \multicolumn{2}{c}{Vendi} & \multicolumn{2}{c}{Distinct bigram} \\
        \cmidrule(lr){3-4}\cmidrule(lr){5-6}\cmidrule(lr){7-8}
        Setting & Norm. & Raw & Norm. & Raw & Norm. & Raw & Norm. \\
        \midrule
        \condHumanOnly{}             & 1.00 & 7.50 & 1.00 & 6.06 & 1.00 & 0.90 & 1.00 \\
        \condAIOnlyLow{}             & 0.33 & 1.83 & 0.24 & 1.82 & 0.31 & 0.44 & 0.49 \\
        \condAIOnlyHigh{}            & 0.95 & 7.17 & 0.96 & 5.46 & 0.93 & 0.91 & 1.02 \\
        \condAIDivHumanConvOutLow{}  & 0.98 & 7.67 & 1.02 & 5.72 & 0.96 & 0.86 & 0.96 \\
        \condAIDivHumanConvOutHigh{} & 1.05 & 8.50 & 1.14 & 6.62 & 1.10 & 0.89 & 0.99 \\
        \bottomrule
    \end{tabular}
\end{table}

\paragraph{Results.} We computed three automatic measures on the same batches and compared them with the mean human cluster counts. The first is $\metricDiversity$, which is used in our main studies (cf.\ Sections~\ref{sec:evaluation1} and~\ref{sec:evaluation2}). The second is the Vendi score, for which we used the same \texttt{Qwen3-Embedding-8B} embeddings with cosine similarity as the kernel and order $q=1$. The third is distinct bigram, which is the number of unique word bigrams divided by the total number of word bigrams in a batch. Across the 10 batches, $\metricDiversity$ correlates with human cluster counts at Spearman $r_s=0.92$ and Pearson $r=0.98$. Vendi has the same Spearman correlation, while distinct bigram has a weaker rank correlation ($r_s=0.39$). Table~\ref{tab:human-rated-diversity} shows that both human judgments and $\metricDiversity$ place \condAIOnlyLow{} well below the human-only baseline, \condAIDivHumanConvOutLow{} near it, and \condAIDivHumanConvOutHigh{} above it. Overall, these results support $\metricDiversity$ as a measure of the semantic distinctions captured by human clustering.

\subsection{Effect of Mixing Models and Personas}
\label{app:model-persona-mixing}

We examine diversity differences associated with persona prompts and model mixing under the policies \textsc{One} and \textsc{Mix}.
We use the same study artifacts analyzed in Sections~\ref{sec:evaluation1} and~\ref{sec:evaluation2}, covering both domains at full adoption ($x=1$).
For each task and interaction mode, we form three groups.
The first group contains the artifacts collected under \textsc{One}, with the AI agents being \texttt{Qwen3-235B-A22B-2507-Instruct} model without a persona.
The second group contains the artifacts collected under \textsc{Mix}, with the AI agents being the same Qwen model with a persona.
The third group contains all artifacts collected under \textsc{Mix}, with AI agents being Qwen model and four other models (cf.\ Section~\ref{subsec:evaluation2-setup}), all with personas. The persona difference compares the diversity of the second group with that of the first, the model mixing difference compares the third group with the second, and the total difference compares the third group with the first. We compare equally sized samples from all groups, using the smallest available group size. We normalize each $\metricDiversity$ difference by the diversity of an equally sized \condHumanOnly{} sample and average over tasks and $1{,}000$ sampling repetitions.

\begin{table}[!htbp]
    \centering
    \setlength{\tabcolsep}{8pt}
    \caption{Mean differences in $\metricDiversity$ at full adoption.
    \textbf{Persona diff.} is the diversity of the Qwen group with personas minus that of the Qwen group without personas.
    \textbf{Model mixing diff.} is the diversity of the full five-model group with personas minus that of its Qwen subset.
    \textbf{Total diff.} is the diversity of the full five-model group with personas minus that of the Qwen group without personas.
    Each difference is normalized by the sampled \condHumanOnly{} diversity and averaged over tasks and repetitions.}
    \label{tab:model-persona-mixing}
    \begin{tabular}{lrrr}
        \toprule
        \multicolumn{4}{c}{\domainHypothesisGeneration{}} \\
        \cmidrule(lr){1-4}
        \textbf{Interaction mode} & \textbf{Persona diff.} & \textbf{Model mixing diff.} & \textbf{Total diff.} \\
        \midrule
        \condAIOnly{}            & $0.238$ & $0.200$ & $0.438$ \\
        \condHumanDivConAIOut{}  & $0.018$ & $0.120$ & $0.138$ \\
        \condAIDivConHumanOut{}  & $0.076$ & $-0.032$ & $0.044$ \\
        \condHumanDivAIConvOut{} & $0.055$ & $0.077$ & $0.132$ \\
        \condAIDivHumanConvOut{} & $0.004$ & $-0.024$ & $-0.020$ \\
        \midrule
        \multicolumn{4}{c}{\domainStoryWriting{}} \\
        \cmidrule(lr){1-4}
        \textbf{Interaction mode} & \textbf{Persona diff.} & \textbf{Model mixing diff.} & \textbf{Total diff.} \\
        \midrule
        \condAIOnly{}            & $0.311$ & $0.151$ & $0.462$ \\
        \condHumanDivConAIOut{}  & $0.072$ & $0.112$ & $0.184$ \\
        \condAIDivConHumanOut{}  & $0.088$ & $0.048$ & $0.135$ \\
        \condHumanDivAIConvOut{} & $0.072$ & $0.112$ & $0.184$ \\
        \condAIDivHumanConvOut{} & $0.006$ & $0.070$ & $0.076$ \\
        \bottomrule
    \end{tabular}
\end{table}

Table~\ref{tab:model-persona-mixing} shows positive total differences in all three interaction modes where AI agent produces the final artifact (\condAIOnly{}, \condHumanDivAIConvOut{}, and \condHumanDivConAIOut{}) in both domains.
Both the persona and model mixing differences are positive in these modes.
Across all five interaction modes, mean persona differences are positive, with the largest in \condAIOnly{} and the next largest in \condAIDivConHumanOut{} in both domains.
Overall, these results support our mixture-based policy as a design choice for promoting diversity.

\clearpage

\section{Study Interfaces and Instructions}
\label{app:instructions}

\subsection{Co-creation Interfaces}
\label{app:cocreation-interfaces}

\begin{figure}[!htbp]
    \centering
    \includegraphics[width=0.95\linewidth]{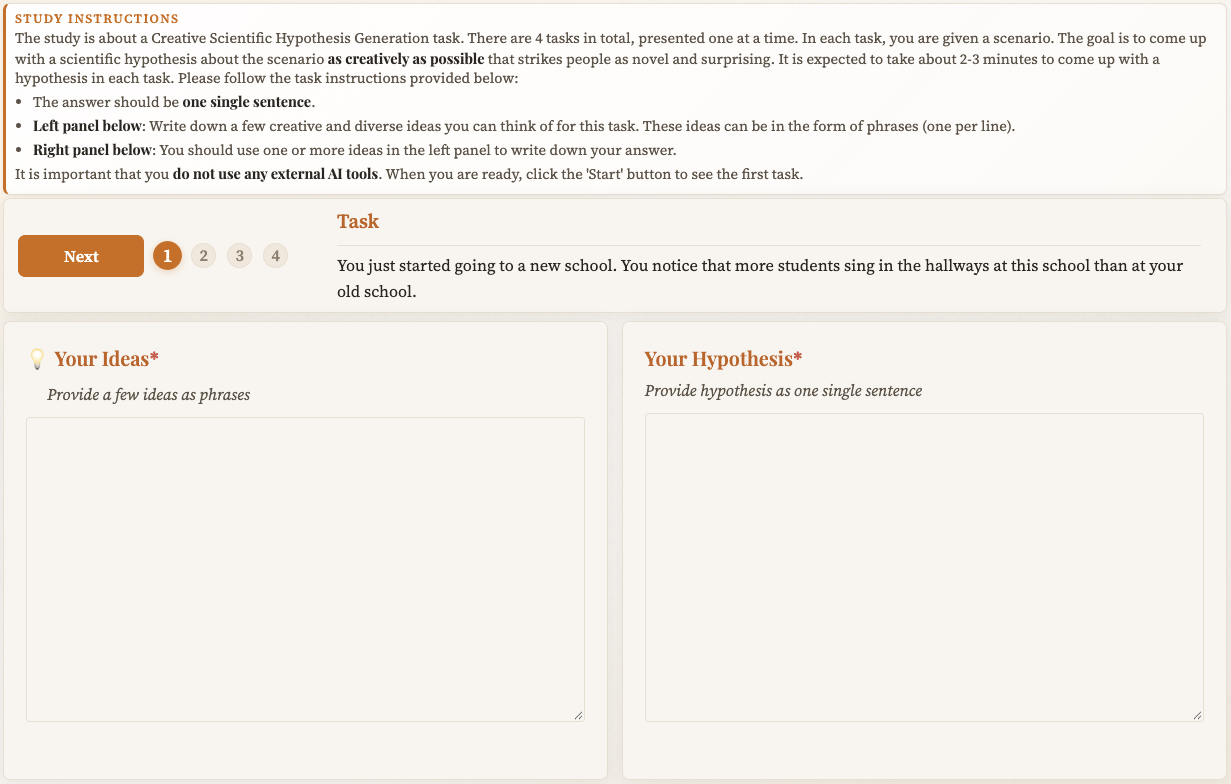}
    \caption{Snapshot of the interface for \condHumanOnly{}.}
    \label{fig:cocreation-snapshot-hhh}
\end{figure}

\begin{figure}[!htbp]
    \centering
    \includegraphics[width=0.95\linewidth]{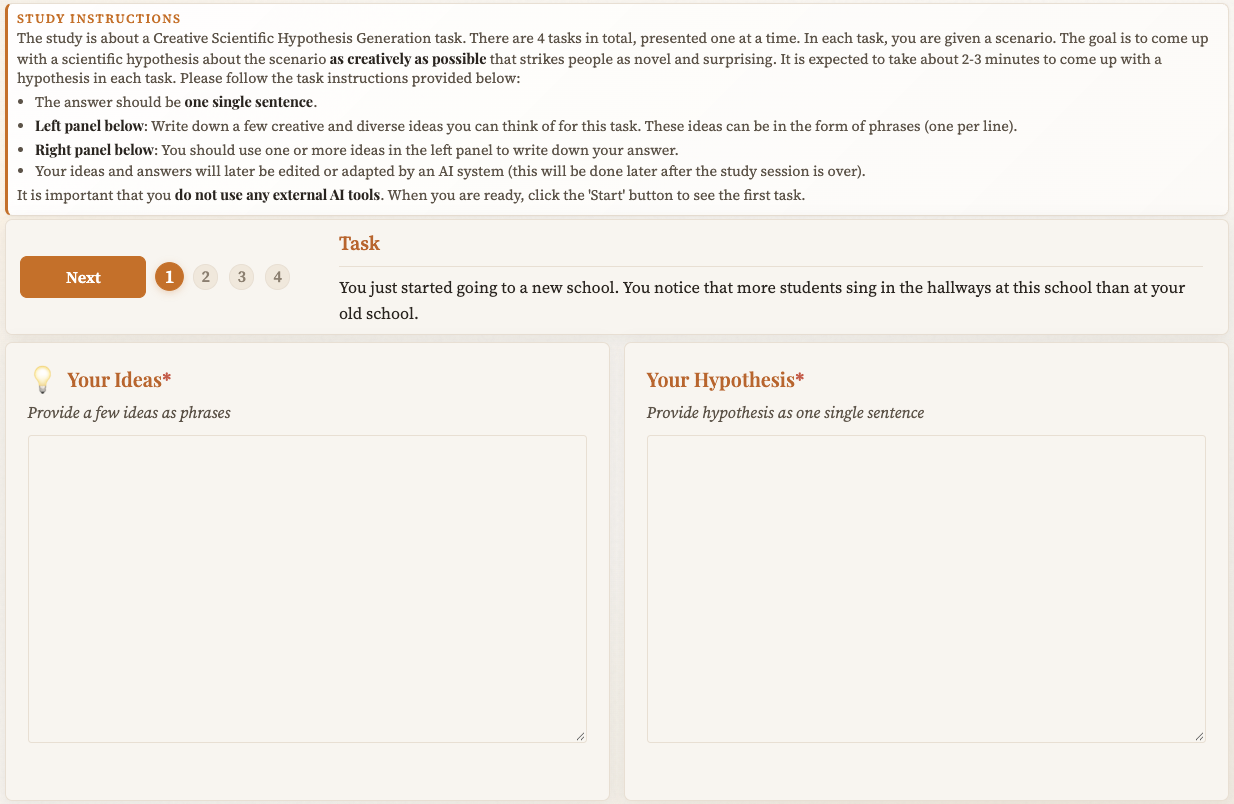}
    \caption{Snapshot of the interface for \condHumanDivConAIOut{}.}
    \label{fig:cocreation-snapshot-hha}
\end{figure}

\begin{figure}[!htbp]
    \centering
    \includegraphics[width=\linewidth]{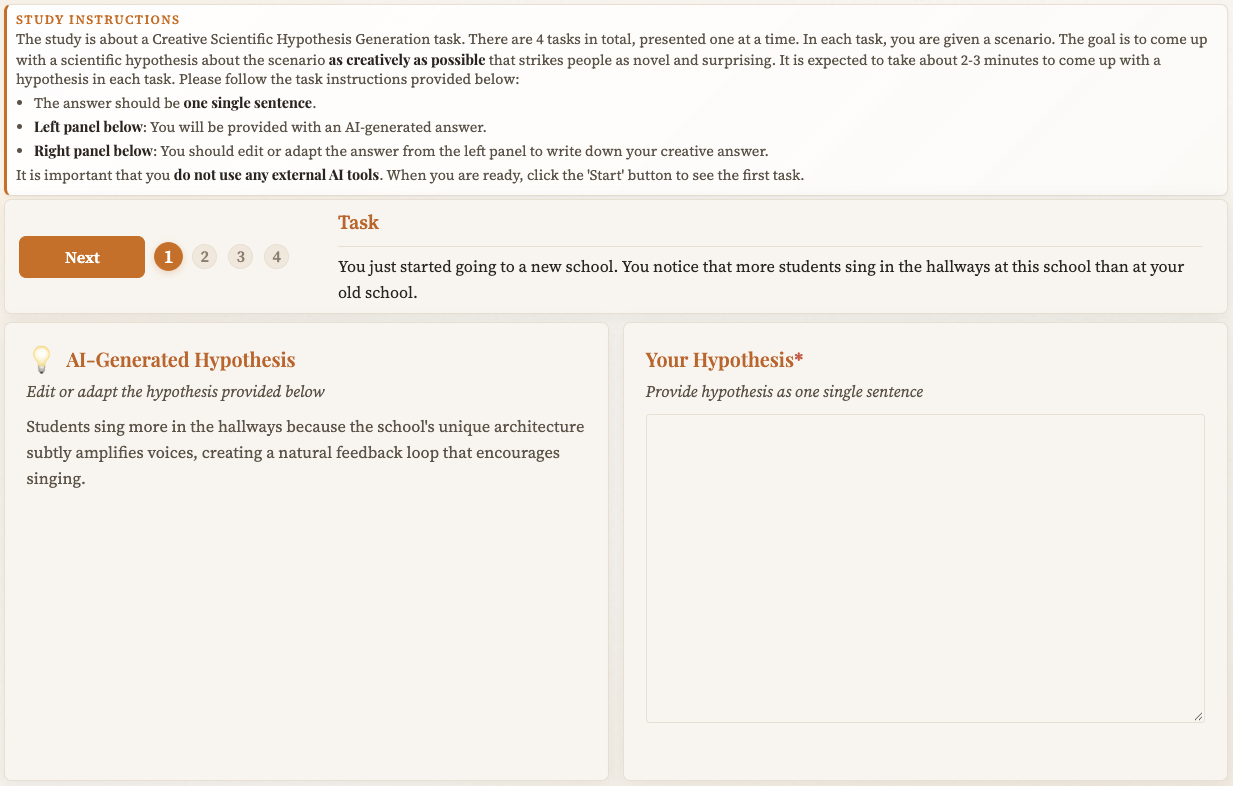}
    \caption{Snapshot of the interface for \condAIDivConHumanOut{}.}
    \label{fig:cocreation-snapshot-aah}
\end{figure}

\begin{figure}[!htbp]
    \centering
    \includegraphics[width=\linewidth]{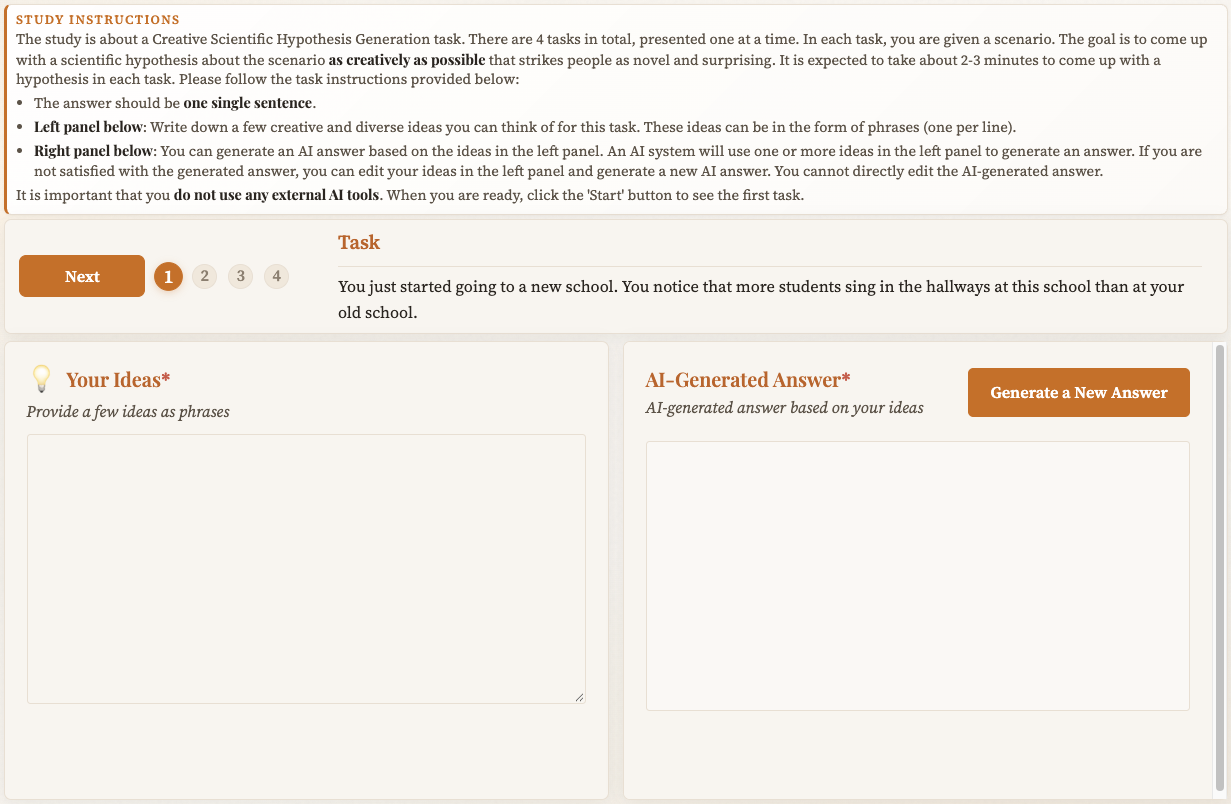}
    \caption{Snapshot of the interface for \condHumanDivAIConvOut{}.}
    \label{fig:cocreation-snapshot-haa}
\end{figure}

\begin{figure}[!htbp]
    \centering
    \includegraphics[width=\linewidth]{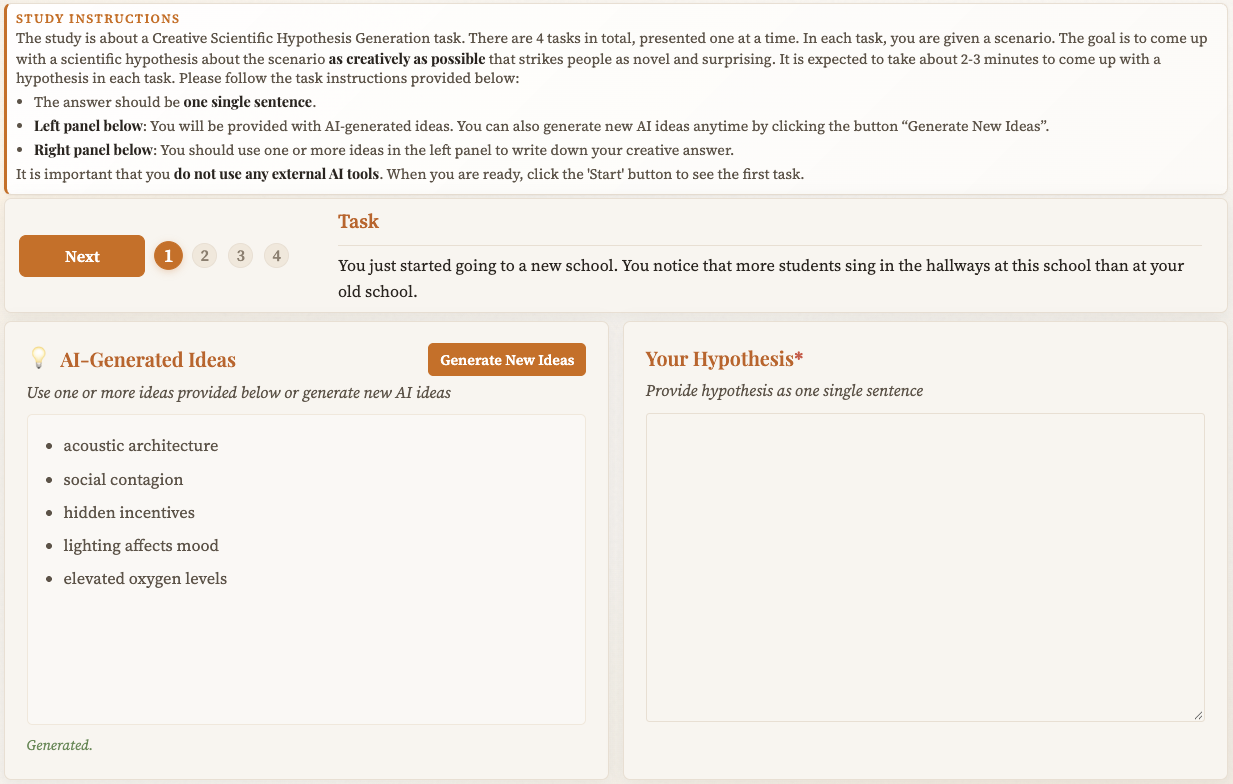}
    \caption{Snapshot of the interface for \condAIDivHumanConvOut{}.}
    \label{fig:cocreation-snapshot-ahh}
\end{figure}

Figures~\ref{fig:cocreation-snapshot-hhh}--\ref{fig:cocreation-snapshot-ahh} show the study interfaces for tasks in \domainHypothesisGeneration{} under the five interaction modes.

\subsection{Annotation Interface}
\label{app:annotation-interface}

\begin{figure}[!htbp]
    \centering
    \includegraphics[width=0.95\textwidth]{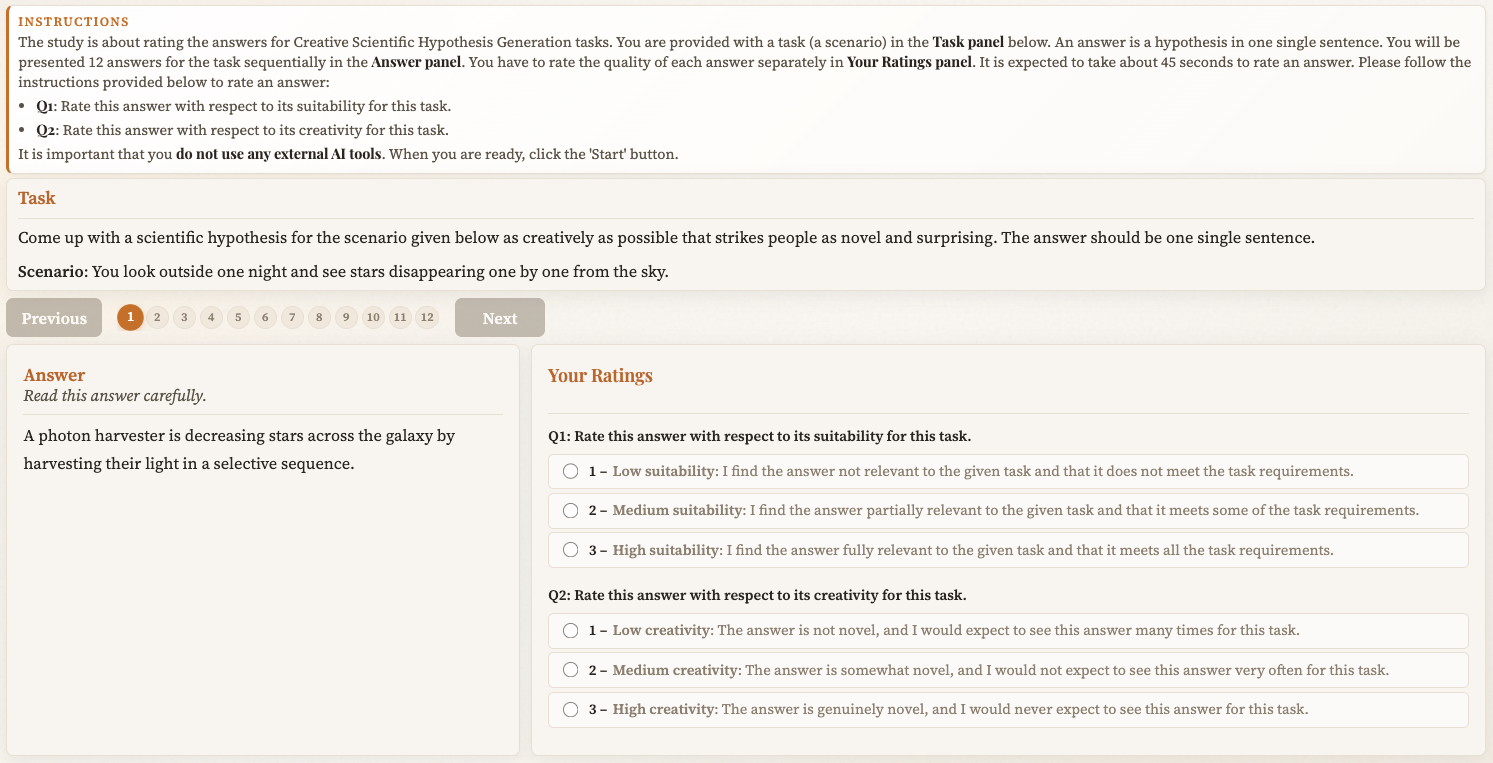}
    \caption{Snapshot of the annotation interface used for human evaluation. Annotators rate the usability (suitability) and creativity of each answer.}
    \label{fig:annotation-snapshot}
\end{figure}

Figure~\ref{fig:annotation-snapshot} shows the annotation interface used for collecting human quality ratings in \domainHypothesisGeneration{}.

\clearpage

\section{Prompts for Agent Responses}
\label{app:prompts}

Figures~\ref{fig:ai-response-prompts-hg} and~\ref{fig:ai-response-prompts-csw} show the prompts used to generate agent responses for \domainHypothesisGeneration{} and \domainStoryWriting{}, respectively.
For the mixture-based policy, the persona placeholder is filled with a persona from PersonaHub~\citep{DBLP:journals/corr/abs-2406-20094}, such as ``A crime novels enthusiast who lives in Washington D.C.''


\begin{figure}[ht!]
    \centering
    \footnotesize
    \newcommand{\promptplaceholder}[1]{\promptinput{\texttt{\{#1\}}}}
    \newcommand{\promptsection}[1]{\textbf{#1}\par\vspace{0.03cm}}

    \begin{subfigure}[t]{\textwidth}
        \centering
        \begin{mdframed}[linewidth=0.5pt,roundcorner=2pt,innertopmargin=6pt,innerbottommargin=6pt,innerleftmargin=7pt,innerrightmargin=7pt]
        \raggedright
        Approach the following task through the perspective of \promptplaceholder{persona}. You will be given a scenario. The goal is to come up with a scientific hypothesis about the scenario as creatively as possible that strikes people as novel and surprising.

        First, think about a few creative and diverse ideas for this task. Each idea is a short phrase of 3--5 words. Then, use one or more ideas to come up with a creative hypothesis. Your answer should be one sentence about your hypothesis. You should avoid using overly complex sentences in your answer. Respond with JSON only in this exact format: \texttt{\{"ideas": ["idea1", "idea2", ...], "answer": "..."\}}.

        \vspace{0.08cm}
        In this task, you are given the following scenario: ``\promptplaceholder{task}''.
        \end{mdframed}
        \caption{Prompt for \condAIDivHumanConvOut{}, \condAIDivConHumanOut{}, and \condAIOnly{}}
        \label{fig:ai-response-prompts-hg-aaa}
    \end{subfigure}

    \vspace{0.8cm}

    \begin{subfigure}[t]{\textwidth}
        \centering
        \begin{mdframed}[linewidth=0.5pt,roundcorner=2pt,innertopmargin=6pt,innerbottommargin=6pt,innerleftmargin=7pt,innerrightmargin=7pt]
        \raggedright
        Approach the following task through the perspective of \promptplaceholder{persona}. You will be given a scenario. The goal is to come up with a scientific hypothesis about the scenario as creatively as possible that strikes people as novel and surprising. You will be given a few creative and diverse ideas for this task. Use one or more of the given ideas to come up with a creative hypothesis. Your answer should be one sentence about your hypothesis. You should avoid using overly complex sentences in your answer. Respond with JSON only in this exact format: \texttt{\{"answer": "..."\}}.

        \vspace{0.08cm}
        In this task, you are given the following scenario and ideas.

        \vspace{0.04cm}
        Scenario: ``\promptplaceholder{task}''

        Ideas: ``\promptplaceholder{human\_ideas}''
        \end{mdframed}
        \caption{Prompt for \condHumanDivAIConvOut{}}
        \label{fig:ai-response-prompts-hg-haa}
    \end{subfigure}

    \vspace{0.8cm}

    \begin{subfigure}[t]{\textwidth}
        \centering
        \begin{mdframed}[linewidth=0.5pt,roundcorner=2pt,innertopmargin=6pt,innerbottommargin=6pt,innerleftmargin=7pt,innerrightmargin=7pt]
        \raggedright
        Approach the following task through the perspective of \promptplaceholder{persona}. You will be given a scenario. The goal is to come up with a scientific hypothesis about the scenario as creatively as possible that strikes people as novel and surprising. You will be given a few creative and diverse ideas together with a hypothesis. Edit or adapt the given hypothesis as you want and use it as your answer. Your answer should be one sentence about your hypothesis. You should avoid using overly complex sentences in your answer. Respond with JSON only in this exact format: \texttt{\{"answer": "..."\}}.

        \vspace{0.08cm}
        In this task, you are given the following scenario, ideas, and hypothesis.

        \vspace{0.04cm}
        Scenario: ``\promptplaceholder{task}''

        Ideas: ``\promptplaceholder{human\_ideas}''

        Hypothesis: ``\promptplaceholder{human\_answer}''
        \end{mdframed}
        \caption{Prompt for \condHumanDivConAIOut{}}
        \label{fig:ai-response-prompts-hg-hha}
    \end{subfigure}

    \caption{Prompts used to generate AI responses for \domainHypothesisGeneration{} in different settings. Placeholders are shown in blue. For each prompt, the first sentence describing the persona is omitted if no persona is specified in the corresponding setting.}
    \label{fig:ai-response-prompts-hg}
\end{figure}


\begin{figure}[ht!]
    \centering
    \footnotesize
    \newcommand{\promptplaceholder}[1]{\promptinput{\texttt{\{#1\}}}}
    \newcommand{\promptsection}[1]{\textbf{#1}\par\vspace{0.03cm}}

    \begin{subfigure}[t]{\textwidth}
        \centering
        \begin{mdframed}[linewidth=0.5pt,roundcorner=2pt,innertopmargin=6pt,innerbottommargin=6pt,innerleftmargin=7pt,innerrightmargin=7pt]
        \raggedright
        Approach the following task through the perspective of \promptplaceholder{persona}. You will be given three words. The goal is to write a story using all three words as creatively as possible that strikes people as novel and surprising.

        First, think about a few creative and diverse ideas for this task. Each idea is a short phrase of 3--5 words. Then, use one or more ideas to come up with a creative story. Your answer should be a story of 3--5 sentences. You should avoid using overly complex sentences in your answer. Respond with JSON only in this exact format: \texttt{\{"ideas": ["idea1", "idea2", ...], "answer": "..."\}}.

        \vspace{0.08cm}
        In this task, you are given the following three words: ``\promptplaceholder{task}''.
        \end{mdframed}
        \caption{Prompt for \condAIDivHumanConvOut{}, \condAIDivConHumanOut{}, and \condAIOnly{}}
        \label{fig:ai-response-prompts-csw-aaa}
    \end{subfigure}

    \vspace{0.8cm}

    \begin{subfigure}[t]{\textwidth}
        \centering
        \begin{mdframed}[linewidth=0.5pt,roundcorner=2pt,innertopmargin=6pt,innerbottommargin=6pt,innerleftmargin=7pt,innerrightmargin=7pt]
        \raggedright
        Approach the following task through the perspective of \promptplaceholder{persona}. You will be given three words. The goal is to write a story using all three words as creatively as possible that strikes people as novel and surprising. You will be given a few creative and diverse ideas for this task. Use one or more of the given ideas to write your story. Your answer should be a story of 3--5 sentences. You should avoid using overly complex sentences in your answer. Respond with JSON only in this exact format: \texttt{\{"answer": "..."\}}.

        \vspace{0.08cm}
        In this task, you are given the following three words and ideas.

        \vspace{0.04cm}
        Three words: ``\promptplaceholder{task}''

        Ideas: ``\promptplaceholder{human\_ideas}''
        \end{mdframed}
        \caption{Prompt for \condHumanDivAIConvOut{}}
        \label{fig:ai-response-prompts-csw-haa}
    \end{subfigure}

    \vspace{0.8cm}

    \begin{subfigure}[t]{\textwidth}
        \centering
        \begin{mdframed}[linewidth=0.5pt,roundcorner=2pt,innertopmargin=6pt,innerbottommargin=6pt,innerleftmargin=7pt,innerrightmargin=7pt]
        \raggedright
        Approach the following task through the perspective of \promptplaceholder{persona}. You will be given three words. The goal is to write a story using all three words as creatively as possible that strikes people as novel and surprising. You will be given a few creative and diverse ideas together with a story written based on the ideas. Edit or adapt the given story as you want and use it as your answer. Your answer should be a story of 3--5 sentences. You should avoid using overly complex sentences in your answer. Respond with JSON only in this exact format: \texttt{\{"answer": "..."\}}.

        \vspace{0.08cm}
        In this task, you are given the following words, ideas, and story.

        \vspace{0.04cm}
        Words: ``\promptplaceholder{task}''

        Ideas: ``\promptplaceholder{human\_ideas}''

        Story: ``\promptplaceholder{human\_answer}''
        \end{mdframed}
        \caption{Prompt for \condHumanDivConAIOut{}}
        \label{fig:ai-response-prompts-csw-hha}
    \end{subfigure}

    \caption{Prompts used to generate AI responses for \domainStoryWriting{} in different settings. Placeholders are shown in blue. For each prompt, the first sentence describing the persona is omitted if no persona is specified in the corresponding setting.}
    \label{fig:ai-response-prompts-csw}
\end{figure}

\clearpage
\section*{NeurIPS Paper Checklist}

\begin{enumerate}

\item {\bf Claims}
    \item[] Question: Do the main claims made in the abstract and introduction accurately reflect the paper's contributions and scope?
    \item[] Answer: \answerYes{} 
    \item[] Justification: The abstract and introduction accurately reflect the paper's contributions and scope, and these claims are supported by Sections~\ref{sec:evaluation1}--\ref{sec:evaluation3}.

\item {\bf Limitations}
    \item[] Question: Does the paper discuss the limitations of the work performed by the authors?
    \item[] Answer: \answerYes{} 
    \item[] Justification: Section~\ref{sec:conclusion} includes the discussion of limitations.

\item {\bf Theory assumptions and proofs}
    \item[] Question: For each theoretical result, does the paper provide the full set of assumptions and a complete (and correct) proof?
    \item[] Answer: \answerNA{} 
    \item[] Justification: We do not include theoretical results in this paper.

    \item {\bf Experimental result reproducibility}
    \item[] Question: Does the paper fully disclose all the information needed to reproduce the main experimental results of the paper to the extent that it affects the main claims and/or conclusions of the paper (regardless of whether the code and data are provided or not)?
    \item[] Answer: \answerYes{} 
    \item[] Justification: Sections~\ref{sec:evaluation1}--\ref{sec:evaluation3} describe the information for the main experimental results. Additional details are provided in the Appendix.

\item {\bf Open access to data and code}
    \item[] Question: Does the paper provide open access to the data and code, with sufficient instructions to faithfully reproduce the main experimental results, as described in supplemental material?
    \item[] Answer: \answerYes{} 
    \item[] Justification: We release our implementation and resources in a Github repository, with a link provided in Section~\ref{sec:introduction}.

\item {\bf Experimental setting/details}
    \item[] Question: Does the paper specify all the training and test details (e.g., data splits, hyperparameters, how they were chosen, type of optimizer) necessary to understand the results?
    \item[] Answer: \answerYes{} 
    \item[] Justification: We specify these experimental details in Sections~\ref{sec:evaluation1}--\ref{sec:evaluation3} and additional details in the Appendix.

\item {\bf Experiment statistical significance}
    \item[] Question: Does the paper report error bars suitably and correctly defined or other appropriate information about the statistical significance of the experiments?
    \item[] Answer: \answerYes{} 
    \item[] Justification: We report 95\% confidence intervals for the main experimental results in Sections~\ref{sec:evaluation1}--\ref{sec:evaluation2}.

\item {\bf Experiments compute resources}
    \item[] Question: For each experiment, does the paper provide sufficient information on the computer resources (type of compute workers, memory, time of execution) needed to reproduce the experiments?
    \item[] Answer: \answerYes{} 
    \item[] Justification: Our experiments primarily involve crowdsourcing studies for which we provide details in Sections~\ref{sec:evaluation1}--\ref{sec:evaluation2} and additional details in the Appendix. AI content is generated through hosted model APIs (Section~\ref{sec:evaluation2}). 
    
\item {\bf Code of ethics}
    \item[] Question: Does the research conducted in the paper conform, in every respect, with the NeurIPS Code of Ethics \url{https://neurips.cc/public/EthicsGuidelines}?
    \item[] Answer: \answerYes{} 
    \item[] Justification: Our studies use consent-based crowdsourcing procedures for data collection and provide appropriate compensation. Our human-subject experiments are designed to conform to the NeurIPS Code of Ethics.

\item {\bf Broader impacts}
    \item[] Question: Does the paper discuss both potential positive societal impacts and negative societal impacts of the work performed?
    \item[] Answer: \answerYes{} 
    \item[] Justification: See Section~\ref{sec:conclusion}.
    
\item {\bf Safeguards}
    \item[] Question: Does the paper describe safeguards that have been put in place for responsible release of data or models that have a high risk for misuse (e.g., pre-trained language models, image generators, or scraped datasets)?
    \item[] Answer: \answerNA{} 
    \item[] Justification: We do not release a foundation model or sensitive dataset.

\item {\bf Licenses for existing assets}
    \item[] Question: Are the creators or original owners of assets (e.g., code, data, models), used in the paper, properly credited and are the license and terms of use explicitly mentioned and properly respected?
    \item[] Answer: \answerYes{} 
    \item[] Justification: The datasets used for our studies are properly cited in Section~\ref{sec:evaluation1}.

\item {\bf New assets}
    \item[] Question: Are new assets introduced in the paper well documented and is the documentation provided alongside the assets?
    \item[] Answer: \answerYes{} 
    \item[] Justification: 
    We provide full documentation with the released assets in a Github repository.

\item {\bf Crowdsourcing and research with human subjects}
    \item[] Question: For crowdsourcing experiments and research with human subjects, does the paper include the full text of instructions given to participants and screenshots, if applicable, as well as details about compensation (if any)? 
    \item[] Answer: \answerYes{} 
    \item[] Justification: We provide details about participant recruitment and compensation in Section~\ref{sec:evaluation1}, and detailed task instructions in Appendix~\ref{app:instructions}.

\item {\bf Institutional review board (IRB) approvals or equivalent for research with human subjects}
    \item[] Question: Does the paper describe potential risks incurred by study participants, whether such risks were disclosed to the subjects, and whether Institutional Review Board (IRB) approvals (or an equivalent approval/review based on the requirements of your country or institution) were obtained?
    \item[] Answer: \answerYes{} 
    \item[] Justification: The study followed a standard protocol for participants recruited from crowdsourcing platforms. The study was approved by a local institutional review board, and all participants provided informed consent (Section~\ref{subsec:evaluation1-setup}). There were no specific risks involved in this study, and the participants were fully informed about the study details.

\item {\bf Declaration of LLM usage}
    \item[] Question: Does the paper describe the usage of LLMs if it is an important, original, or non-standard component of the core methods in this research? Note that if the LLM is used only for writing, editing, or formatting purposes and does \emph{not} impact the core methodology, scientific rigor, or originality of the research, declaration is not required.
    \item[] Answer: \answerYes{} 
    \item[] Justification: LLMs serve as AI agents in our co-creation studies. We describe the models, agent policies, and prompts in Sections~\ref{sec:evaluation1}--\ref{sec:evaluation2} and Appendix~\ref{app:prompts}.

\end{enumerate}

\end{document}